\documentclass[12pt, draftclsnofoot, onecolumn]{IEEEtran}
\normalsize

\usepackage{enumerate}
\usepackage{color}

\usepackage{cite}

\usepackage{float}
\newfloat{figtab}{htb}{fgtb}
\makeatletter
\newcommand\figcaption{\def\@captype{figure}\caption}
\newcommand\tabcaption{\def\@captype{table}\caption}
\makeatother

\usepackage{graphicx}
\ifCLASSINFOpdf
\else
\fi

\usepackage{amsthm}
\usepackage{amssymb,amsfonts}
\usepackage{amsfonts,balance}
\usepackage{bbm}
\usepackage{dsfont}

\usepackage{bm}

\usepackage[cmex10]{amsmath}

\newtheorem{theorem}{Theorem}

\newtheorem{remark}{Remark}

\newtheorem{lemma}{\textbf{Lemma}}
\usepackage{booktabs}

\usepackage[ruled,linesnumbered]{algorithm2e}

\usepackage{array}
\usepackage{multirow,makecell}
\usepackage[caption=false,font=footnotesize]{subfig}

\usepackage{verbatim}

\begin{document}
%
\title{Microfluidic QCSK Transmitter and Receiver Design for Molecular Communication}
%
%
%

\author{Dadi~Bi,~\IEEEmembership{Student Member,~IEEE,}
        Yansha~Deng,~\IEEEmembership{Member,~IEEE,}\\
\thanks{D. Bi and Y. Deng are with the Department of Engineering, King’s College London, London, WC2R 2LS, U.K. (e-mail:\{dadi.bi, yansha.deng\}@kcl.ac.uk). (Corresponding author: Yansha Deng).}
}
\maketitle

\vspace{-1.8cm}
\begin{abstract}
The design of components with molecular communication (MC) functionalities can bring an opportunity to enable some emerging applications in fields from personal healthcare to modern industry. In this paper, we propose the designs of the microfluidic transmitter and receiver with quadruple concentration shift keying (QCSK) modulation and demodulation functionalities. To do so, we first present an AND gate design, and then apply it to the QCSK transmitter and receiver design. The QCSK transmitter is capable of modulating two input signals to four different concentration levels, and the QCSK receiver can demodulate a received signal to two outputs. More importantly, we also establish a mathematical framework to theoretically characterize our proposed microfluidic circuits. Based on this, we first derive the output concentration distribution of our proposed AND gate design, and provide the insight into the selection of design parameters to ensure an exhibition of desired behaviour. We further derive the output concentration distributions of the QCSK transmitter and receiver. Simulation results obtained in COMSOL Multiphysics not only show the desired behaviour of all the proposed microfluidic circuits, but also demonstrate the accuracy of the proposed mathematical framework. 
\end{abstract}


\vspace{-0.6cm}

%
\IEEEpeerreviewmaketitle

\vspace{-0.9cm}	
\section{Introduction}
Over the past few years, molecular communication (MC) has attracted increasing attention as it can wave revolutionary and interdisciplinary applications ranging from healthcare, to industry, and military \cite{8726417,8742793}. This inspired a bulk of research centering around theoretical characterizations of MC, such as transmission schemes \cite{5962989,7397863}, propagation characterizations \cite{6807659,twoway}, reception models \cite{Yansha16,8030318}, and detection strategies \cite{6712164,8922790}. 
To ensure successful information transmission, signal processing units are envisioned to be essential components for MC transmitter and receiver to facilitate modulation/demodulation and coding/decoding functionalities. However, how to practically realize these basic signal processing functions in microscale/nanoscale has been rarely studied. 

The signal processing functions realized in existing MC works were performed over electrical signals using electrical devices. In \cite{7397863,giannoukos2018chemical,8489889,farsad2013tabletop,8924625}, the transmitted bit sequence was modulated over the concentration of signalling molecules via the on/off of an air tank \cite{7397863,giannoukos2018chemical,8489889}, electrical spray \cite{farsad2013tabletop}, and LED controlled by Arduino microcontroller boards and laptops \cite{8924625}. Their high dependency over electrical signals/devices can hardly fulfil the biocompatible and non-invasive requirements of biomedical applications, such as disease diagnosis and drug delivery \cite{akyildiz2008nanonetworks,andreescu2004trends}. Meanwhile, the size of electrical devices can hardly meet the requirement of intra-body healthcare applications promised by MC, where fully MC functional devices are expected to be miniaturized into microscale/nanoscale \cite{8726417}.

In nature, signal processing functions can be realized over molecular domain by exploiting the gene expression process,
where transcription factors bind with genes to either activate or repress their expression into proteins \cite{Alberts2009}. A gene expression process can be functioned as a buffer gate if the transcription factor activates the protein expression \cite{wang2014rapid}, and can be functioned as a NOT gate if a gene expression is repressed by the transcription factor \cite{Wang11}. This signal processing nature motivates biologists to 
design more complex computing artificial genetic circuits to manipulate molecular concentrations using the synthetic biology \cite{kahl2013survey}. One type of artificial genetic circuits with computing functions is the Boolean logic inspired digital logic devices. The sharp state change {between a low and high concentration} is ideal for reliable state transitions and signal integration, making digital logic particularly useful in decision-making circuits \cite{xiang2018scaling}. For example, the authors in \cite{Wang11} designed an orthogonal AND gate and coupled it to nonspecific sensors to increase selectivity \cite{bernard2017synthetic}. The authors in \cite{tamsir2011robust} constructed a simple NOR logic gate and spatially configured multiple NOR gates to produce all possible two-input gates, which have found their utilities in biotechnological applications \cite{ellis2009diversity}.

Although the aforementioned genetic circuits has advantages in biocompatibility and miniaturization over electrical circuits, the experimental realization of genetic circuits for signal processing faces challenges, such as slow speed, unreliability, and non-scalability \cite{alon2006introduction}. This motivates our initial work on chemical reactions-based microfluidic circuits \cite{8255057,bi2019chemical,dadimag}. {Unlike genetic circuits, chemical circuits are much easier to be controlled, and their integration with microfluidic devices brings advantages in low reagents consumption, rapid analysis, and high efficiency \cite{whitesides2006origins}.} In \cite{8255057,bi2019chemical}, we designed an MC microfluidic transceiver based on chemical reactions to successfully realize the binary concentration shift keying (BCSK) modulation and demodulation functions. The signal processing capability of chemical reactions-based microfluidic circuits is further exploited in \cite{dadimag}, where we provided the designs of the AND gate, NAND gate, NOR gate, OR gate, and XOR gate, which are validated through COMSOL simulations. Echoing the discussion in \cite{dadimag}, one challenge in realizing signal processing functions via chemical reactions-based microfluidic circuits design is the theoretical characterization of the logic gate, which facilitates the design parameters selection for the expected gate outputs. Although we mathematically modelled the dynamics of
molecular species in microfluidic channels in \cite{8255057,bi2019chemical}, this analysis is not scalable with the increase in the number of microfluidic circuits. 
Motivated by above, the main contributions of this paper are as follows:
\begin{itemize}
	\item We first present a chemical reactions-based microfluidic AND gate design, based on which, we design the microfluidic transmitter and receiver with quadruple concentration shift keying (QCSK) modulation and demodulation functionalities, to show how logic computations can process molecular concentrations. Compared with the BCSK transceiver in \cite{bi2019chemical}, the QCSK transceiver can achieve two bits signal transmission with improved data rate.	
	\item We develop a novel mathematical framework to characterize our proposed microfluidic circuits, which can be applied to analyse other new and more complicated microfluidic circuits. To do so, we first analyse the concentration and velocity changes under fluid convergence and separation, and we derive the transfer function of a straight convection-diffusion channel. Based on these, we derive the spatial-temporal concentration distribution of a convection-diffusion-reaction channel with either a thresholding reaction or an amplifying reaction.
	\item To evaluate our proposed microfluidic designs, we identify {four} elementary microfluidic blocks of the basic AND gate, and define five corresponding operators to represent the output concentration distribution of each elementary block. Relying on this, we not only derive the output concentration distribution of the proposed AND gate, but also those for our designed QCSK transmitter and receiver. The functionalities of our proposed microfluidic designs and the corresponding theoretical results are validated via simulations performed in COMSOL Multiphysics finite element solver.
\end{itemize}

The remainder of this paper is organized as follows. In Sec. \ref{sec:basic_analysis}, we provide the basic microfluidic channel analysis. In Sec. \ref{sec:AND}, we establish a mathematical framework to theoretically characterize our proposed AND gate. In Sec. \ref{sec:QCSK}, we propose the designs and analysis of the QCSK transmitter and receiver. Numerical results in Sec. \ref{sec:simulation} validate the proposed microfluidic designs and their theoretical analysis. Finally. Sec. \ref{sec:con} concludes the paper.

\vspace{-0.5cm}
\section{Basic Microfluidic Channel Analysis}
\label{sec:basic_analysis}
\vspace{-0.2cm}
With the ultimate goal of designing and analysing a microfluidic system with modulation and demodulation functionalities in this paper, the basic characteristics of fluids in microfluidic channels must be first understood. To do so, we first analyse the concentration and velocity changes due to the hydrodynamic fluid convergence in Sec. \ref{fluid_convergence} and the fluid separation in Sec. \ref{fluid_separation}. We then present the molecular concentration distribution of a convection-diffusion channel in Sec. \ref{convection_diffusion}.


For a Poiseuille flow travelling along the $x$ direction of a microfluidic channel with rectangular cross-section, the velocity profile can be obtained by solving the Navier-Stokes equation, which is \cite{bruustheoretical}
\vspace{-0.2cm}
\begin{align}
v^{\text{}}_{}(y,z)=\frac{4h^2\Delta p}{\pi^3\eta L} \sum_{n,\text{odd}}^{\infty} \frac{1}{n^3}[1-\frac{\text{cosh}(\frac{n\pi y}{h})}{\text{cosh}(\frac{n\pi w}{2h})} ]\text{sin}(\frac{n\pi z}{h}), \label{velocity1}
\end{align}
where ${\Delta p}/{L}$ denotes the pressure difference between two ends of a channel with length $L$, $\eta$ is the fluid dynamic viscosity, and $w$ and $h$ are the width and the height of the cross-section. 

The fluid velocity follows a ``parabolic" distribution, where the closer to the channel centre, the larger the fluid velocity, resulting in the maximum velocity $v_\text{max}$ occurring at the centre of the channel. Alternatively, it is common to use the average velocity $v_{\text{eff}}$ to describe the fluid velocity, which can be presented as
\vspace{-0.3cm}
\begin{align}
	v_{\text{eff}}=\frac{Q}{wh}.
	\label{velocity3}
\end{align}
In \eqref{velocity3}, $Q$ is the volumetric flow rate, which represents the fluid volume that passes per unit time, and can be calculated as the integral of velocity contributions from each lamina using \cite{bruustheoretical}
\begin{equation}
	\begin{aligned}
	Q^\text{}&=\int_S v^{\text{}}_{}(y,z)\mathrm{d}y \mathrm{d}z \\
	&=\frac{8h^3w\Delta p}{\pi ^3\eta L}\sum_{n=1,3,5,\cdots}^{\infty}[\frac{1}{n^4}-\frac{2h}{\pi w n^5}\text{tanh}(\frac{n\pi w}{2h})].
	\end{aligned} 
	\label{velocity4}
\end{equation}

\vspace{-0.45cm}
\subsection{Fluid Convergence at Combining Connections}
\label{fluid_convergence}
\vspace{-0.2cm}
In a microfluidic circuit, fluids flowing in different channels can converge to a single flow at a combining connection, and we name this behaviour as fluid convergence for simplicity.

\vspace{-0.35cm}
\subsubsection{Concentration Change}
\label{sec:variation_analysis}
\begin{figure}[!t]
	\centering
	\includegraphics[width=3.8in]{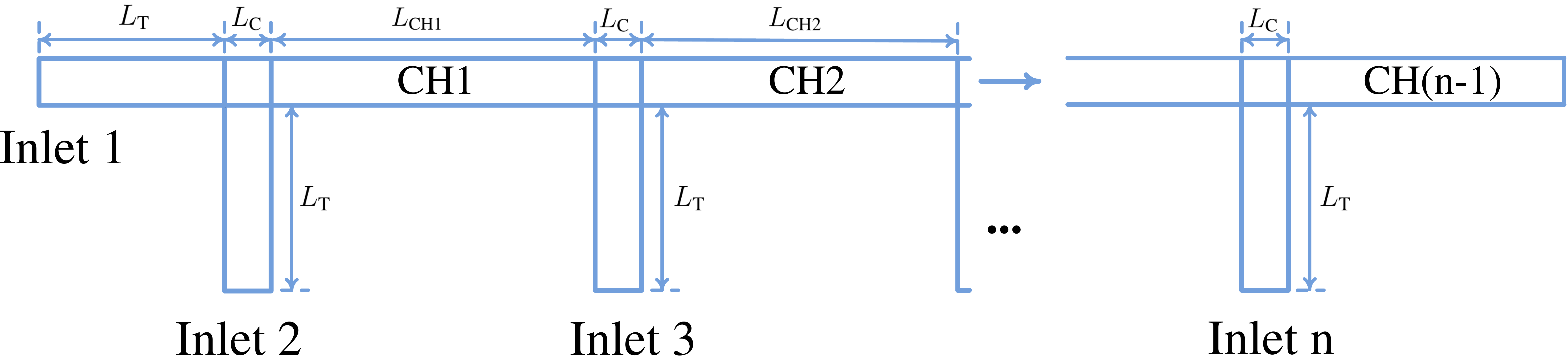}
	\caption{A microfluidic device for fluid convergence analysis.}
	\label{f_microfluidic_device}
\end{figure}
Let us consider a microfluidic device with $n$ inlets and $n-1$ combining channels as shown in Fig. \ref{f_microfluidic_device}. We assume that 
species $S_i$ $(1\leq i \leq n)$ is constantly injected into Inlet $i$ with concentration $C_{S_{i_0}}$, average velocity $v_{\text{eff}_i}$, and volumetric flow rate $Q_i$. 
According to the well-known analogy between hydraulic circuits and electrical circuits, the pressure drop, the flow rate, and the flow resistance in hydraulic circuits are analogous to the voltage drop, the electrical current, and the electrical resistance in electrical circuits, respectively. Based on the \textit{Kirchhoff’s Current Law}, the volumetric flow rate in the $i$th combining channel $Q^{\text{CH}i}$ is the sum of coming flow rates, such that 
\begin{equation} \label{KCL}
\left\{
	\begin{array}{lr}
		Q^{\text{CH}1}=Q_{1}+Q_{2}, \\
		Q^{\text{CH}2}=Q^{\text{CH}1}+Q_{3},\\
		~~~~~~~~~~\cdots \cdots,\\
		Q^{\text{CH}(n-1)}=Q^{\text{CH}(n-2)}+Q_{n}.
	\end{array}
\right.
\end{equation}
Therefore, the mixed concentrations of species $S_1$ and $S_2$ in the first combining channel are \cite{oh2012design}
\begin{equation}
	\left\{
		\begin{array}{lr}
			C^{\text{CH}1}_{S_1}&=\frac{Q_1}{Q_1+Q_2} C_{S_{1_0}},\\
			C^{\text{CH}1}_{S_2}&=\frac{Q_2}{Q_1+Q_2} C_{S_{2_0}}.
		\end{array}
	\right.
\end{equation}
Similarly, the mixed concentrations of species $S_1$, $S_2$, $\cdots$, $S_i$ in the $(i-1)$th combining channel become
\begin{equation} \label{mixed_concentration}
\left\{
\begin{array}{lr}
	C^{\text{CH}(i-1)}_{S_1}=\frac{Q_1+Q_2+\cdots+Q_{i-1}}{(Q_1+Q_2+\cdots+Q_{i-1})+Q_i}C^{\text{CH}(i-2)}_{S_1},\\
	C^{\text{CH}(i-1)}_{S_2}=\frac{Q_1+Q_2+\cdots+Q_{i-1}}{(Q_1+Q_2+\cdots+Q_{i-1})+Q_i}C^{\text{CH}(i-2)}_{S_2},\\
	~~~~~~~~~~~~~~~~~ \cdots \cdots~~~~~~~~~~,\\
	C^{\text{CH}(i-1)}_{S_i}=\frac{Q_i}{(Q_1+Q_2+\cdots+Q_{i-1})+Q_i}C_{S_{i_0}}.
\end{array}
\right.
\end{equation}
\begin{lemma}
\label{lemma_combining_concentration}
For the fluid convergence from $n$ inlets to one combining channel, the mixed concentration of species $S_i$ can be derived as
	\begin{align} \label{lemma1_eq1}
	C^{{\rm{CH}}(i-1)}_{S_i}=\frac{Q_i}{{\sum_{j=1}^{i}} Q_j}C_{S_{i_0}},
	\end{align}
where $Q_i$ and $C_{S_{i_0}}$ are the volumetric flow rate and the species concentration injected into Inlet $i$. If all the species are injected with volumetric flow rate $Q$, \textit{i.e.}, $Q_1=\cdots=Q_n=Q$, species $S_i$ will be diluted to $1/i$ of its injected concentration in the $(i-1)$th combining channel, that is
	\begin{align} \label{combining_concentration}
	C^{{\rm{CH}}(i-1)}_{S_i}=\frac{1}{i}C_{S_{i_0}}.
	\end{align}
\end{lemma}
\vspace{-0.5cm}
\begin{proof}
	The last line of \eqref{mixed_concentration} can be easily reduced to \eqref{lemma1_eq1}.
\end{proof}
\vspace{-0.5cm}
\begin{remark}
From \eqref{combining_concentration}, we can conclude that the more inlet channels with the same volumetric flow rate injected, the lower the concentration is.
\end{remark}

%

\vspace{-0.4cm}
\subsubsection{Velocity Change}
\label{sec:velocity_analysis}
Injecting fluids into a combining channel influence not only the species concentration but also the velocity profile.
\vspace{-0.4cm} 
\begin{lemma}
	\label{lemma_combining_velocity}
	For the fluid convergence from $n$ inlets to one combining channel, the flow rate in the combining channel can be expressed in terms of average velocity and channel geometry as
	\begin{align} 
	w^{{\rm{CH}}{(i-1)}}h^{{\rm{CH}}{(i-1)}}v^{{\rm{CH}}{(i-1)}}_{{\rm{eff}}}=
	w^{{\rm{CH}}(i-2)}h^{{\rm{CH}}(i-2)}v^{{\rm{CH}}(i-2)}_{\rm{eff}}+w_{i}h_{i}v_{{\rm{eff}}_{i}}, \label{v_variation1}
	\end{align}
	where $v_{{\rm{eff}}_{i}}$, $w_{i}$, and $h_{i}$ are the average velocity, the width, and the height of Inlet $i$, and $v^{{\rm{CH}}i}_{\rm{eff}}$, $w^{{\rm{CH}}i}$, and $h^{{\rm{CH}}i}$ are the average velocity, the width, and the height of the $i$th combining channel, respectively. 
	If all inlets and combining channels share the same geometries and the same average velocity $v_{\rm{eff}}$, the average velocity in the $(i-1)$th combining channel becomes
	\begin{align} \label{combining_velocity}
	v^{{\rm{CH}}(i-1)}_{\rm{eff}}=iv_{\rm{eff}}.
	\end{align}
\end{lemma}
\vspace{-0.5cm}
\begin{proof}
	Based on the \textit{Kirchhoff’s Current Law} and \eqref{velocity3}, we can obtain \eqref{v_variation1}.
\end{proof}
\vspace{-0.5cm}
\begin{remark}
	It is revealed in \eqref{combining_velocity} that the more inlet channels with the same volumetric flow rate injected, the larger the average velocity is.
\end{remark}

\vspace{-0.7cm}
\subsection{Fluid Separation at Bifurcation Connections}
\label{fluid_separation}
\begin{figure*}[!t]
	\centering
	\subfloat[{A microfluidic device}\label{bifurcation_microfluidic}]{\includegraphics[width=2.2in]{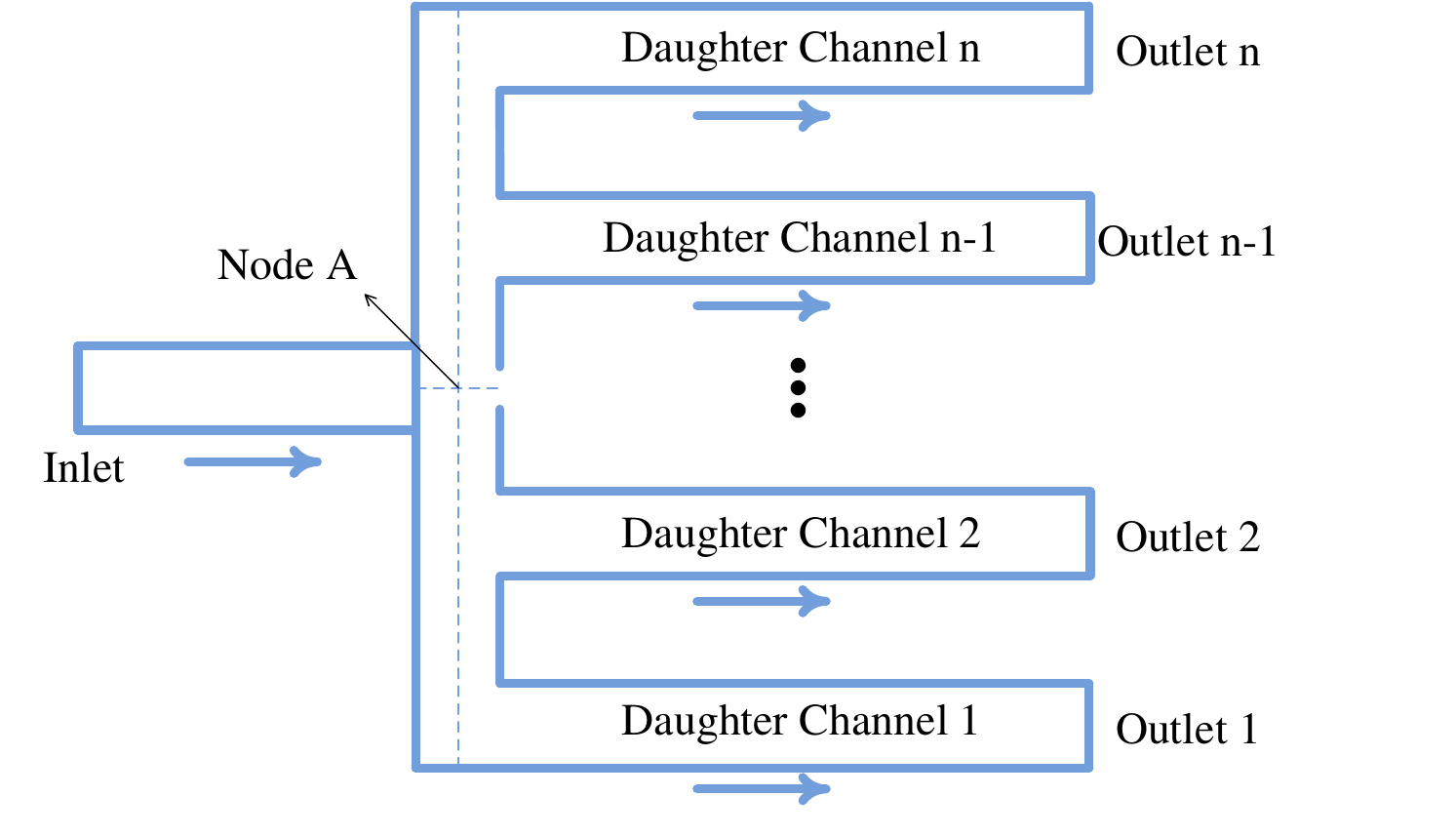}}
	\subfloat[Hydraulic circuit analog\label{bifurcation_elec}]{\includegraphics[width=2.2in]{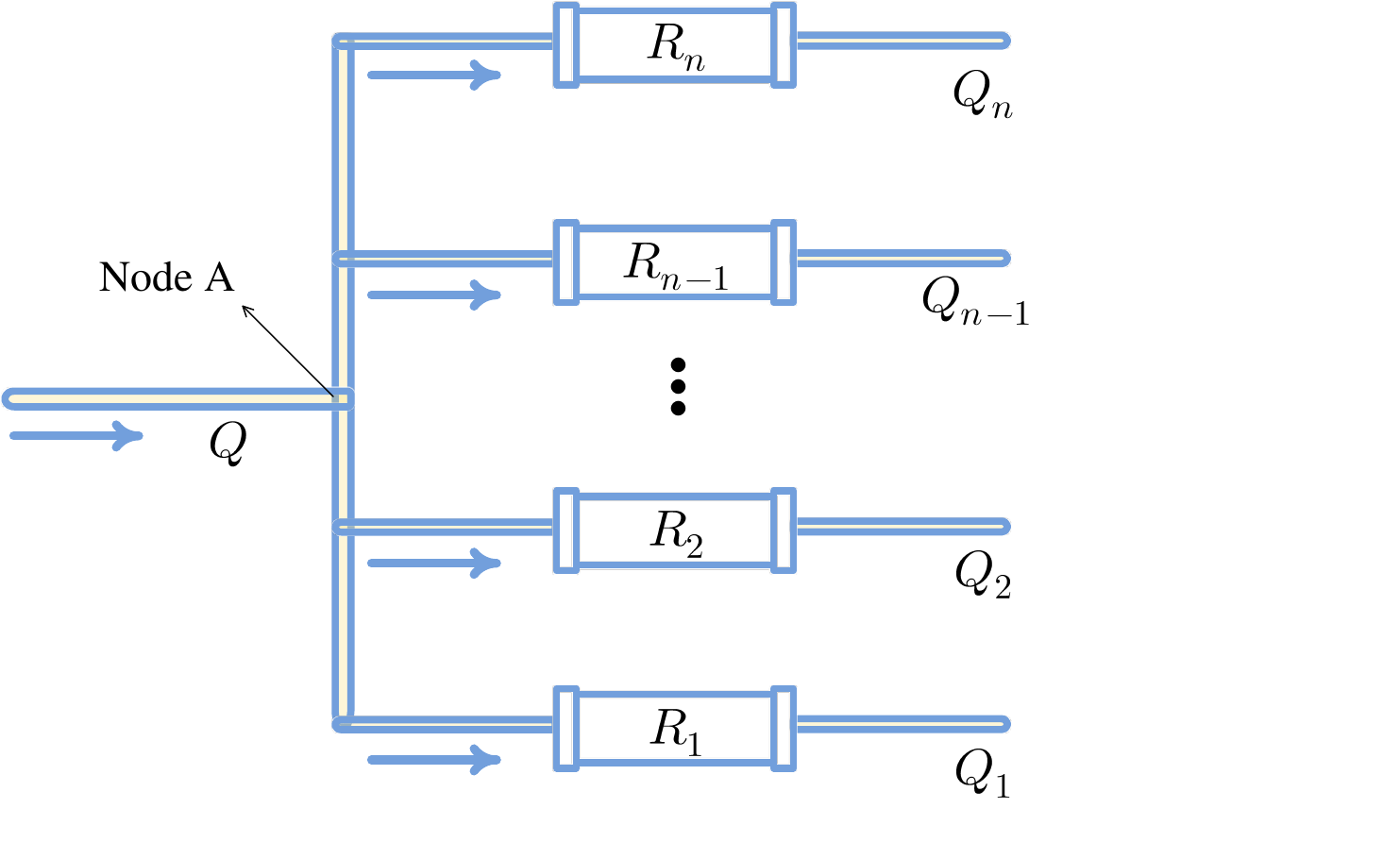}}
	\caption{A microfluidic device for fluid separation analysis.}
	\label{fig:bifurcation}%
\end{figure*}

In a microfluidic circuit, a single flow can be separated into different flow streams at a bifurcation connection, and we name this behaviour as fluid separation for simplicity. Let us consider a microfluidic device with one inlet and $n$ outlets as shown in Fig. \ref{fig:bifurcation}\subref{bifurcation_microfluidic}, where a single flow is separated into $n$ streams travelling over $n$ daughter channels.
Assuming that species $S_1$ is injected with concentration $C_{S_{1_0}}$ and average velocity $v_{\text{eff}}$, the concentration at each outlet is the same as $C_{S_{1_0}}$, because species $S_1$ is not diluted by other species. However, the average velocity in each outlet varies for different geometries of its daughter channel. To derive the outlet average velocities, we establish the hydraulic circuit model in Fig. \ref{fig:bifurcation}\subref{bifurcation_elec}. Analogous to current division in electric circuits, the relationship between the volumetric flow rate $Q_i$ $(1\le i \le n)$ and the supplied volumetric flow rate $Q$ can be described by \cite[Eq. (18)]{oh2012design}
\begin{align} \label{Q_division}
Q_i=\frac{R_{eq}}{R_i}Q,
\end{align}    
where $R_i$ is the hydraulic resistance of the $i$th daughter channel and $R_{eq}$ is the equivalent resistance of all daughter channels. Let us denote $L_{D_i}$ as the length from the crosspoint Node A in Fig. \ref{fig:bifurcation}\subref{bifurcation_elec} to outlet $i$, and $w_i$ and $h_i$ as the geometry width and height of the $i$th daughter channel, $R_i$ \cite[Eq. (6)]{toh2014engineering} and $R_{eq}$ can be calculated as 
\begin{align}
	&R_i=\frac{12\eta L_{D_i}}{w_i h_i^3 \biggl( 1-\sum\limits_{i=1,3,5,\cdots}^{\infty} \frac{192h_i}{w\pi^5 i^5}\text{tan}h_i(\frac{i\pi h_i}{2w_i}) \biggr)},\label{resistance_cal} \\
	\text{and}~~&R_{eq}={1}/({{1}/{R_1}+{1}/{R_2}+\cdots+{1}/{R_n}}). \label{eq_resistance}
\end{align}
\begin{lemma}
	\label{lemma_bifurcation_velocity}
	For the fluid separation from one inlet to $n$ outlets, the average velocity $v_{{\rm{eff}}_i}$ in the $i$th outlet can be derived as
	\begin{align} \label{v_division}
	v_{{\rm{eff}}_i}=\frac{1}{L_{D_i}\sum_{i=1}^{n}\frac{1}{L_{D_i}}}\frac{wh}{w_i h_i} v_{\rm{eff}}.
	\end{align}
	\vspace{-0.1cm}If all daughter channels share the same geometries, \eqref{v_division} can be reduced to
	\begin{align} \label{v_division_reduce}
	v_{{\rm{eff}}_i}=\frac{1}{n}v_{\rm{eff}}.
	\end{align}
\end{lemma}
\vspace{-0.45cm}
\begin{proof}
	Combining \eqref{velocity3}, \eqref{Q_division}-\eqref{eq_resistance}, we can obtain \eqref{v_division}.
\end{proof}
\vspace{-0.5cm}
\begin{remark}
	It is indicated from \eqref{v_division_reduce} that fluid separation results in a reduction of average velocity by $n$ times.
\end{remark}

\vspace{-1.1cm}
\subsection{Convection-Diffusion Channel}
\vspace{-0.2cm}
\label{convection_diffusion}
When a flow containing species $S_i$ enters the dispersion regime \cite{wicke2018modeling}, we can describe the spatial-temporal concentration distribution of species $S_i$ using a 1D convection-diffusion equation as 
\begin{align}
\frac{\partial C_{S_i}(x,t)}{{\partial}t}=D_{\text{eff}}\frac{\partial^2 C_{S_i}(x,t)}{{\partial}x^2}-v_{\text{eff}}\frac{\partial C_{S_i}(x,t)}{{\partial}x}, \label{cd}
\end{align} 
where 
$D_{\text{eff}}$ is the \textit{Taylor-Aris} effective diffusion coefficient. For a microfluidic channel with rectangular-shaped cross-section whose height is $h$ and width is $w$, $D_\text{eff}$ can be calculated as  \cite{bicen2014end}
\vspace{-0.3cm}
\begin{align}
	D_{\text{eff}}=1+\frac{8.5{v_{\text{eff}}^2}{h^2}{w^2}}{210D^2({h^2}+2.4hw+{w^2})},
\end{align}
where $D$ is the molecular diffusion coefficient.

Although the solution of \eqref{cd} with a rectangular input has been derived in \cite{bi2019chemical}, its complex expression does not allow the cascaded channels to be mathematically solvable in closed-form. This motivates us to derive the transfer function of a microfluidic channel so that the output of a microfluidic circuit can be written as the convolution of an input and a cascade of the transfer function of each channel. We solve the transfer function in the following theorem.
\vspace{-0.3cm}
\begin{theorem}
	\label{theo_cd}
	The transfer function of a straight convection-diffusion channel is derived as
	\begin{equation} \label{approx2}
	\begin{aligned}
	{H_{\text{}}}(x,t)=\frac{1}{2 \pi}\int_0^\infty [e^{-j\omega t}\overline{\widetilde{{C_{S_i}}}(x,\omega)}+e^{j\omega t}{\widetilde{{C_{S_i}}}(x,\omega)}]\mathrm{d}w,
	\end{aligned}
	\end{equation}	
	where $	\widetilde{{C_{S_i}}}(x,s)=\exp\big[  	{{{v_{\rm{eff}}x}/{2D_{\rm{eff}}}}{-\sqrt{x^2({v_{\rm{eff}}^2+4jw{D_{\rm{eff}}})}/{4{D_{\rm{eff}}}^2}}}} \big].$
\end{theorem}
\vspace{-0.5cm}
\begin{proof}
	See the Appendix \ref{app_cd}.
\end{proof}

\vspace{-0.2cm}
From \textbf{Theorem 1}, the outlet concentration of species $S_i$ can be expressed as
\begin{align}
	C_{S_i}(x,t)=C_{S_{i_0}}(t)*H(x,t), \label{theo1}
\end{align}
where $C_{S_{i_0}}(t)$ is the input concentration of species $S_i$ at channel inlet and ``$*$'' denotes the convolution operator.

\vspace{-0.6cm}
\section{AND Logic Gate Design and Analysis}
\label{sec:AND}
In this section, we present the design of the AND logic gate to demonstrate the logic computation ability of microfluidic circuits. The chemical reactions used in the AND gate can be categorized into two forms: 1) the thresholding reaction $S_i+S_j \to S_k$, and 2) the amplifying reaction $S_i+Amp\to S_i+O$. To characterize the outlet concentration of our designed AND gate, we first study the concentration distribution of the reaction channel with either a thresholding reaction, or an amplifying reaction. Relying on the analysis in Sec. \ref{sec:basic_analysis}, we then define and model {four} elementary blocks in order to theoretically characterize the AND gate.

\begin{figure}
	\centering
	\includegraphics[width=5in]{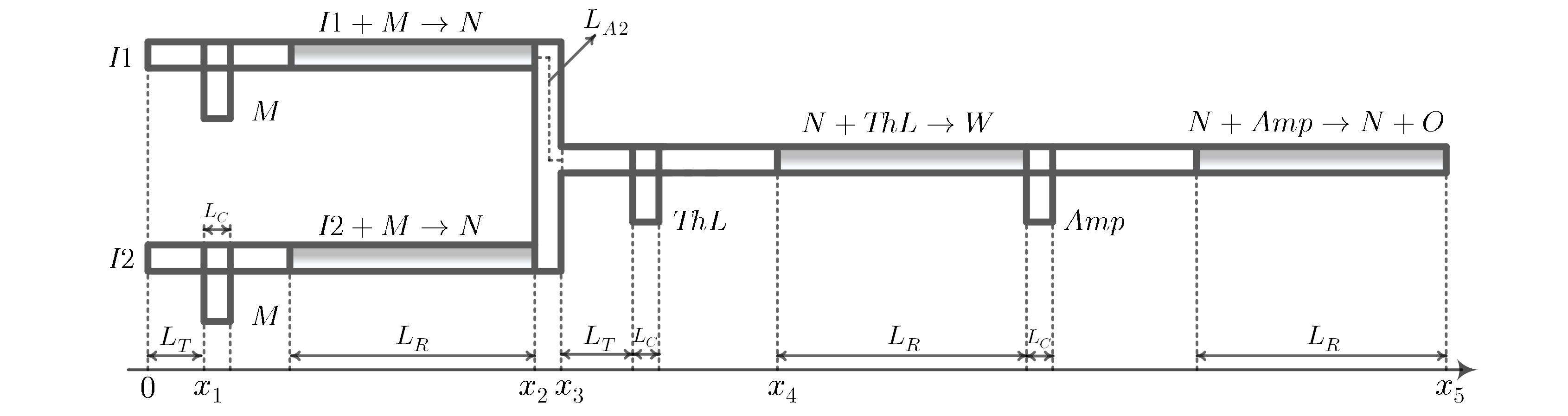}
	\caption{The chemical reactions-based microfluidic AND logic gate. To distinguish convection-diffusion channels and convection-diffusion-reaction channels, the latter are filled with grey-gradient colour.}
	\label{AND_gate_design}
\end{figure}

The proposed \textbf{AND gate design} is presented in Fig. \ref{AND_gate_design}.
As shown in Fig. \ref{AND_gate_design}, the proposed AND gate consists of the input species $I1$ and $I2$, and the output species $O$. {Throughout this paper, we use non-zero concentration to represent HIGH state (bit-1), and zero concentration to represent LOW state (bit-0).} The two input species $I1$ and $I2$ are first converted to an intermediate species $N$, so the concentration of species $N$ will be HIGH state if either $I1$ or $I2$ is HIGH. Then, species $N$ is further depleted by species $ThL$ to extract the interval where both the inputs are HIGH. Finally, the remaining species $N$ catalyses the conversion of species $Amp$ to output species $O$.

\vspace{-0.6cm}
\subsection{Channel Model with a Thresholding Reaction}
\label{sec:thresholding}
\vspace{-0.2cm}
For a straight microfluidic channel with the thresholding reaction $S_i +S_j \to S_k$, the spatial-temporal concentration distributions of reactant and product can be expressed by convection-diffusion-reaction equations, which are expressed as
\begin{align}
\frac{\partial C_{S_{i}}(x,t)}{{\partial}t}=D_{\text{eff}}\frac{\partial^2 C_{S_{i}}(x,t)}{{\partial}x^2}-v_{\text{eff}}\frac{\partial C_{S_{i}}(x,t)}{{\partial}x}-kC_{S_i}(x,t)C_{S_j}(x,t), \label{cdr1} \\
\frac{\partial C_{S_k}(x,t)}{{\partial}t}=D_{\text{eff}}\frac{\partial^2 C_{S_k}(x,t)}{{\partial}x^2}-v_{\text{eff}}\frac{\partial C_{S_k}(x,t)}{{\partial}x}+kC_{S_i}(x,t)C_{S_j}(x,t) \label{cdr3},
\end{align}
where $k$ is the rate constant. Compared with the convection-diffusion equation in \eqref{cd}, the newly introduced reaction term is fully coupled with convection and diffusion process, which complicates the resolution of \eqref{cdr1} and \eqref{cdr3}. A strategy to tackle this coupling is to apply the ``operator splitting" method. It first separates an original differential equation into several parts, then separately computes the solution of each part, and finally combines these separate solutions to form a solution for the original equation. As we already derived the transfer function of a straight convection-diffusion channel in \textbf{Theorem 1}, this motivates us to separate a convection-diffusion-reaction equation into a reaction term and a convection-diffusion term. This separation can be achieved via 1) assuming the reactants are added into a virtual reactor, and the unconsumed reactants and generated product flow into a convection-diffusion channel as soon as the reaction stops; and 2) treating the solution of the reaction term as the initial input for the convection-diffusion part.


{With species $S_i$ and $S_j$ continuously flowing into a channel, we regard that $S_i$ and $S_j$ are continuously added into a virtual reactor, where the continuous reactant supply is a superposition of reactant addition with constants at different times.} To solve this, we consider the following two scenarios:
\begin{itemize}
	\item \textbf{Scenario 1}: species $S_{i}$ and $S_j$ are only added at $t=0$ with concentration $C_{S_{i_0}}$ and $C_{S_{j_0}}$,
	\item \textbf{Scenario 2}: species $S_{i}$ and $S_j$ are added continuously with concentration $C_{S_{i_0}}(t)$ and $C_{S_{j_0}}(t)$.
\end{itemize}
{We first derive concentration changes of reactants and product for \textbf{Scenario 1}, which will then be applied in \textbf{Scenario 2} to derive the solutions of the separated reaction term.}

\subsubsection{\textbf{Scenario 1}}
Let $c(t)$ denote the consumed concentration of reactant $S_i/S_j$ during the reaction. It is noted that $c(t)$ can also represent the concentration of generated species $S_k$ during the reaction due to a one-to-one stoichiometric relation between reactants and product. The remaining concentrations of species $S_{i}$ and $S_j$ can be expressed as 
\vspace{-0.2cm}
\begin{subequations}
	\begin{alignat}{2}
		C_{S_{i}}(t)=C_{{S_{i_0}}}-c(t), \\
		C_{S_j}(t)=C_{{S_j}_0}-c(t).
	\end{alignat}
\end{subequations}
Then, the reaction equation can be expressed as \cite[Eq. (9.13)]{chang2005physical}
\begin{align}
\frac{\mathrm{d}[C_{{S_{i_0}}}-c(t)]}{\mathrm{d} t}=-k[C_{{S_{i_0}}}-c(t)][C_{{S_{j_0}}}-c(t)]. \label{rate_law1}
\end{align}
After rearrangement, \eqref{rate_law1} becomes
\begin{align}
\frac{\mathrm{d}c(t)}{[C_{{S_{i_0}}}-c(t)][C_{{S_{j_0}}}-c(t)]}=k\mathrm{d}t. \label{rate_law2}
\end{align}
By taking the integral of the two sides of \eqref{rate_law2}, we yield 
\begin{equation}
c(t)=\begin{cases}
\frac{C_{{S_{i_0}}}C_{{S_{j_0}}}\exp{[ (C_{{S_{j_0}}}-C_{{S_{i_0}}})kt]}-C_{{S_{i_0}}}C_{{S_{j_0}}}}{C_{{S_{j_0}}}\exp{[(C_{{S_{j_0}}}-C_{{S_{i_0}}})kt]}-C_{{S_{i_0}}}}, ~~C_{{S_{i_0}}}\le C_{{S_{j_0}}},\\
\frac{C_{{S_{i_0}}}C_{{S_{j_0}}}\exp{[ (C_{{S_{i_0}}}-C_{{S_{j_0}}})kt]}-C_{{S_{i_0}}}C_{{S_{j_0}}}}{C_{{S_{i_0}}}\exp{[(C_{{S_{i_0}}}-C_{{S_{j_0}}})kt]}-C_{{S_{j_0}}}}, ~~C_{{S_{i_0}}}\ge C_{{S_{j_0}}} .  
\end{cases}
\label{rate_law3}
\end{equation}

\begin{remark}
	It can be observed from \eqref{rate_law3} that $c(t)$ is proportional to the rate constant $k$. The higher the rate constant, the faster a reactant is consumed and decreased to zero.
\end{remark}
\vspace{-0.7cm}
\begin{lemma} \label{lemma_scenario1}
	 When the reaction rate $k \to \infty$,
	 the consumed concentration $c(t)$ of reactant $S_i/S_j$ with thresholding reaction $S_i+S_j\to S_k$ in the microfluidic channel can be derived as
	 \vspace{-0.3cm}
	\begin{align}
	\lim_{k\to\infty} 
	c(t)=\varphi(C_{{S_{i_0}}},C_{{S_{j_0}}}), \label{rate_law6}
	\end{align}
	where $C_{S_{i_0}}$ and $C_{S_{j_0}}$ are the injected concentrations of specie $S_i$ and $S_j$, and	$\varphi (\cdot,\cdot)$ is defined as
	\vspace{-0.5cm}
	\begin{align} \label{f_max}
	\varphi(x,y)=\min{\{x,y\}}.	
	\end{align}
\end{lemma}
\vspace{-0.3cm}
\begin{proof} 
	With $k \to \infty$, \eqref{rate_law3} can be easily reduced to \eqref{rate_law6}.
\end{proof}

\subsubsection{\textbf{Scenario 2}}
We consider the continuous injection of species $S_{i}$ and $S_j$ with concentration $C_{S_{i_0}}(t)$ and $C_{S_{j_0}}(t)$. \textbf{Scenario 2} can be regarded as a superposition of \textbf{Scenario 1} in time domain. To apply the analysis of \textbf{Scenario 1}, we first discretize the reaction process into many time intervals with the step $\Delta t$. Thus, the added concentration of species $S_i$ can be denoted as $C_{{S_{i},a}}^n=C_{{S_{i_0}}}(n\Delta t)$ $(n\ge 0)$, where the subscript $a$ refers to addition. We also denote $C_{{S_{i_0}}}^n$ and $C_{{S_i,r}}^n$ as the initial and the remaining concentrations of $S_i$ at $t=n\Delta t$, respectively. The same notations are also applied to species $S_j$.

\begin{algorithm}[t]
	\caption{The Calculation of Remaining Concentrations of Species $S_i$ and $S_j$}
	\label{Algorithm}
	\KwIn{The input concentrations $C_{{S_{i_0}}}(t)$ and $C_{{S_{j_0}}}(t)$. The calculation time interval $[0, T]$. The time step $\Delta t$.}
	Initialization of $C_{{S_{i_0}}}^0$$=$$C_{{S_i,a}}^0$ and $C_{{S_{j_0}}}^0$$=$$C_{{S_j,a}}^0$. \
	
	\For{$n\gets 1, \lfloor {{T/\Delta t}} \rfloor $}{
		
		Calculate the consumed concentration $c^{n-1}$ during $[(n-1)\Delta t, n\Delta t]$ according to \eqref{rate_law3} by interchanging $C_{S_{i_0}}$$\to$$ C_{{S_{i_0}}}^{n-1}$ and $C_{S_{j_0}}$$\to$$C_{{S_{j_0}}}^{n-1}$. \
		
		Update the remaining concentration $C_{S_i,r}^n$$=$$C_{S_{i_0}}^{n-1}$$-$$c^{n-1}$ and $C_{S_j,r}^n$$=$$C_{S_{j_0}}^{n-1}$$-$$c^{n-1}$.\
		
		Update the initial concentration $C_{S_{i_0}}^n$$=$$C_{S_{i},r}^n$$+$$C_{S_{i},a}^n$ and $C_{S_{j_0}}^n$$=$$C_{S_{j},r}^n$$+$$C_{S_{j},a}^n$ for $[n\Delta t, (n+1)\Delta t]$.\
		
	}
\end{algorithm}
We propose \textbf{Algorithm \ref{Algorithm}} to numerically calculate the remaining concentrations of $S_i$ and $S_j$ after reaction $S_i +S_j \to S_k$. \textbf{Algorithm \ref{Algorithm}} describes that for any time interval $[n\Delta t, (n+1)\Delta t]$, the consumed concentration can be calculated according to \eqref{rate_law1}, but with different initial concentrations $C_{{S_{i_0}}}^n$. This is due to the fact that the initial concentration at any time interval is not only influenced by the newly added concentration, but also the incompletely consumed concentration that added in previous intervals. For instance, the initial concentration $C_{{S_{i_0}}}^1$ for the time interval $[\Delta t, 2\Delta t]$ is the sum of the newly added concentration $C_{{S_{i},a}}^1$ and the remaining unreacted concentration $C_{{S_{i},r}}^1$ that added at $t=0$.

\begin{figure}
	\centering
	\includegraphics[width=5.6in]{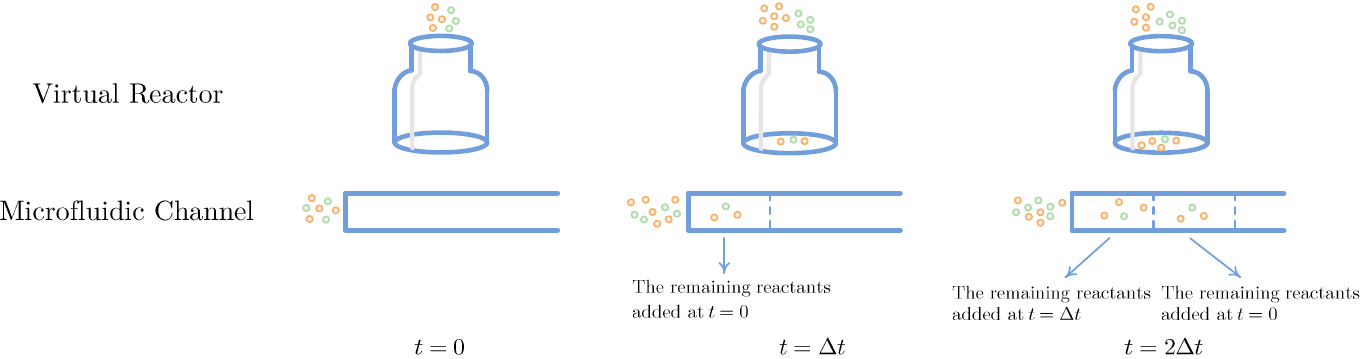}
	\caption{Illustration of reaction $S_i +S_j \to S_k$ in a bottle-shaped virtual reactor and a microfluidic channel. The two reactants are marked with different colours.}
	\label{fig:reactor}%
\end{figure}

It is noted that the value of the rate constant $k$ influences the accuracy of the approximation. The smaller the $k$, the larger volume of reactants remain. The unconsumed reactants accumulate in reactor and would participate into the reaction of the next time interval, which introduces correlation between different time intervals. By contrast, this correlation does not exist in practical scenario. As shown in Fig. \ref{fig:reactor}, 
for time interval $[\Delta t, 2\Delta t]$, the flowing fluid carries remaining reactants added at $t=0$ and $t=\Delta t$ forward, which prevents them from interacting with each other. {Therefore, in the virtual reactor, we should make the rate constant approach infinity to ensure reaction is always completed inside any time interval, thus eliminating the correlation.}  
With $\Delta t \to 0$, the remaining concentrations of $S_i$ and $S_j$ calculated in \textbf{Algorithm 1} reduce to
\begin{subequations}
	\begin{alignat}{2}
	C_{{S_{i}},r}(t)=C_{{S_{i_0}}}(t)&-\varphi [C_{{S_{i_0}}}(t), C_{{S_{j_0}}}(t)],\label{surplus_i} \\
	C_{{S_{j}},r}(t)=C_{{S_{j_0}}}(t)&-\varphi [C_{{S_{i_0}}}(t), C_{{S_{j_0}}}(t)] \label{surplus_j},
	\end{alignat}
\end{subequations}
where $\varphi(\cdot,\cdot)$ is given in \eqref{f_max}.

We derive the output concentration distributions of species $S_i$, $S_j$ and $S_k$ in the following lemma.
\vspace{-0.4cm}
\begin{lemma} \label{lemma_thresholding}
	For a straight reaction channel with the thresholding reaction $S_i+S_j\to S_k$, the output concentration distributions of species $S_i$, $S_j$, and $S_k$ can be derived as
	\begin{subequations} 
		\begin{alignat}{2} 
		C_{S_{i}}(x,t)= C_{S_{i},r}(t)*H(x,t), \label{solution_cdr1}\\
		C_{S_j}(x,t)= C_{S_j,r}(t)*H(x,t) \label{solution_cdr2},\\
		C_{S_k}(x,t)= \varphi [C_{{S_{i_0}}}(t), C_{{S_{j_0}}}(t)]*H(x,t),
		\label{solution_cdr3}
		\end{alignat}
	\end{subequations}
where $C_{S_{i},r}(t)$, $C_{S_{j},r}(t)$, $H(x,t)$, and $\varphi(\cdot,\cdot)$ are given in \eqref{surplus_i}, \eqref{surplus_j}, \eqref{approx2}, and \eqref{f_max}, respectively.
\end{lemma}
\vspace{-0.4cm}
\begin{proof}
	Recall that we separate a convection-diffusion-reaction equation into a reaction part and a convection-diffusion part, we consider the remaining concentrations of $S_i$ in \eqref{surplus_i} and $S_j$ in \eqref{surplus_j} as inputs to a straight convection-diffusion channel. According to {\eqref{theo1}}, we can obtain \eqref{solution_cdr1} and \eqref{solution_cdr2}. The derivation of \eqref{solution_cdr3} can see Appendix B.
\end{proof}

\vspace{-0.79cm}
\subsection{Channel Model with an Amplifying Reaction}
\label{sec:amplifying}
\vspace{-0.2cm}
\begin{lemma} \label{lemma_amplification}
	For a straight reaction channel with the amplifying reaction $S_i+Amp\to S_i+O$, the output concentration distribution of species $O$ can be derived as
	\begin{align}
	C_{O}(x,t)= \big[ C_{Amp_0}(t)\cdot \mathbbm{1}_{\{C_{S_{i_0}}(t)>0\} } \big]*H(x,t), \label{amp_approximation}
	\end{align}
	where $C_{Amp_0}(t)$ and $C_{S_{i_0}}(t)$ are the injected concentrations of species $Amp$ and $S_i$, $u(t)$ is the Heaviside step function, $\mathbbm{1}_{\{\cdot\}}$ is the indicator function that represents the value 1 if the statement is true, and zero otherwise.
\end{lemma}
\vspace{-0.55cm}
\begin{proof} 
	To analyse a straight microfluidic channel with amplifying reaction $S_i +Amp \to S_i+O$, we also separate it into a reaction term and a convection-diffusion term. For the reaction term, as species $O$ is only produced in the presence of $S_i$ and the concentration equals the injected concentration of species $Amp$ \cite{scalise2014designing}, the reaction solution can be expressed as $C_{Amp_0}(t)\cdot\mathbbm{1}_{\{C_{S_{i_0}}(t)>0\} }$.  
	Taking the reaction solution as the initial input for a convection-diffusion channel, the concentration of product $O$ can be derived as \eqref{amp_approximation}. 
\end{proof}

\vspace{-0.7cm}
\subsection{Elementary Blocks}
\vspace{-0.2cm}
Relying on the analysis of fluid convergence in \textbf{Lemma \ref{lemma_combining_concentration} and \ref{lemma_combining_velocity}}, convection-diffusion channel in \textbf{Theorem 1}, and convection-diffusion-reaction channel in \textbf{Lemma \ref{lemma_thresholding} and \ref{lemma_amplification}}, we focus on the analysis of {four} elementary blocks in our designed AND gate (Fig. \ref{AND_gate_design}) in Table \ref{operator}. Meanwhile, we define five typical operators for the {four} elementary blocks, aiming at simplifying the output expression of a microfluidic circuit.
\begin{table*}[tb]
	\centering
	\caption{{Four} elementary blocks.}
	{\renewcommand{\arraystretch}{0.9}
		\scalebox{0.85}
		{\begin{tabular}{c |c |c }
				\hline
				\hline
				Operator & Elementary Block  & Operator Output \\ \hline
				\makecell{$\mathcal{T}[C_{S_{i_0}}(t),n]$\\ Eq. \eqref{operator_T}} & \begin{minipage}{0.57\textwidth}
					\centering
					\includegraphics[width=3.2in]{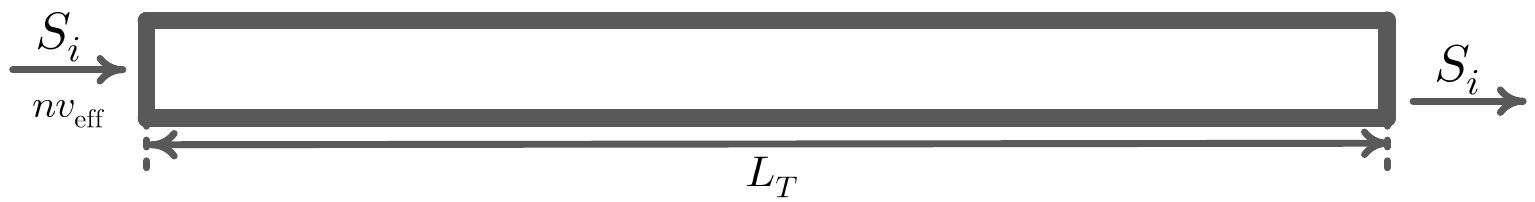}
				\end{minipage} & \makecell{$C_{S_i}(t)$: The output of a\\ convection-diffusion channel\\ with length $L_T$.} \\ \hline
				\makecell{$\mathcal{G}[C_{S_{i_0}}(t),C_{S_{j_0}}(t),n]$\\ Eq. \eqref{operator_G}} & \multirow{2}*{	 \begin{minipage}{0.57\textwidth}
						\centering
						\includegraphics[width=3.4in]{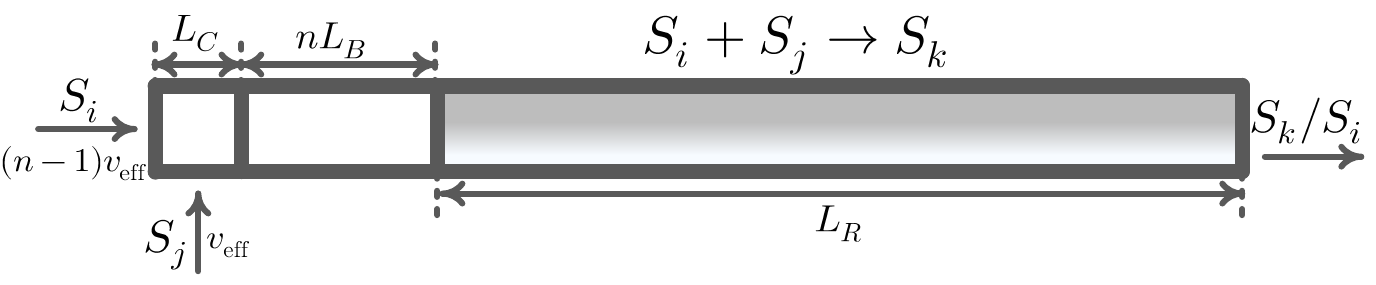}
				\end{minipage}} & \makecell{$C_{S_k}(t)$: The concentration of\\ product $S_k$ with $S_i +S_j \to S_k$.} \\ \cline{1-1} \cline{3-3}
				\makecell{$\mathcal{R}[C_{S_{i_0}}(t),C_{S_{j_0}}(t),n]$\\ Eq. \eqref{operator_R}} &
				& \makecell{$C_{S_i}(t)$: The remaining concentration \\of $S_i$ with $S_i +S_j \to S_k$.} \\
				\hline
				\makecell{$\mathcal{A}[C_{S_{i_0}}(t),C_{Amp_0}(t),n]$\\ Eq. \eqref{operator_A}} & \begin{minipage}{0.57\textwidth}
					\centering
					\includegraphics[width=3.4in]{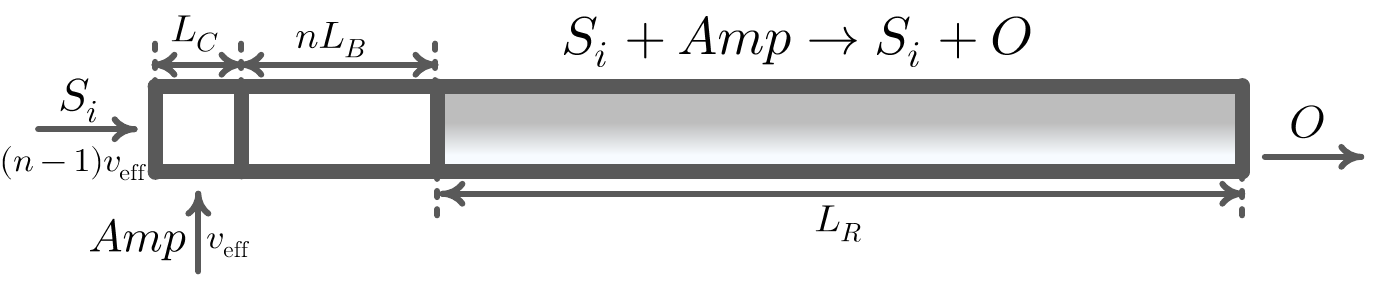}
				\end{minipage} & \makecell{$C_{O}(t)$: The concentration of \\product $O$ with \\$S_i +Amp \to S_i+O$.} \\
				\hline
				\makecell{$\mathcal{F}[C_{S_i}(t),C_{S_j}(t),C_{Amp_0}(t),n]$\\ Eq. \eqref{operator_F}} & \begin{minipage}{0.57\textwidth}
					\centering
					\includegraphics[width=3.8in]{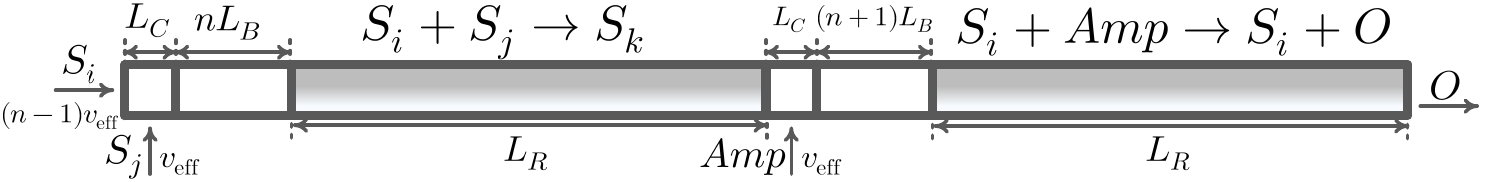}
				\end{minipage} & \makecell{$C_{O}(t)$: The concentration of \\product $O$ with $S_i +S_j \to S_k$\\ and $S_i +Amp \to S_i+O$.} \\
				\hline
				\hline
	\end{tabular}}}
	\label{operator}
\end{table*} 
As shown in Table \ref{operator}, the operator $\mathcal{T}[\cdot]$ represents the output of a convection-diffusion channel with length $L_T$, and can be expressed as
\begin{align}
\mathcal{T}[C_{S_{i_0}}(t),n]\triangleq C_{S_{i_0}}(t)*H_{n}(L_T,t), \label{operator_T}
\end{align}
where the subscript $n$ of $H_n$ indicates that the average velocity in the channel is $nv_\text{eff}$.

For the block with thresholding reaction $S_i +S_j \to S_k$, species $S_i$ with concentration $C_{S_{i_0}}(t)$ and velocity $(n-1)v_\text{eff}$, and species $S_j$ with concentration $C_{S_{j_0}}(t)$ and velocity $v_\text{eff}$,
are injected to the channel with length $L_C$. The convergence of two subchannels with species $S_i$ and $S_j$ will result in a concentration dilution, and the diluted concentrations of $S_i$ and $S_j$ are ${(n-1)C_{{S_{i_0}}}(t)}/{n}$ and ${C_{{S_{j_0}}}(t)}/{n}$ following \eqref{lemma1_eq1} in \textbf{Lemma \ref{lemma_combining_concentration}}, respectively.
Meanwhile, the average velocity will increase to $nv_\text{eff}$ following {\eqref{v_variation1}} in \textbf{Lemma \ref{lemma_combining_velocity}}. Then, species will flow to a buffer channel before the convection-diffusion-reaction channel filled  with  grey-gradient  colour. The buffer channel allows the reactants to be well mixed before a reaction, and the reactant mixing along the radial direction only relies on diffusion. To achieve a fully diffusional mixing, the minimum buffer length $L_B$ can be estimated as
\vspace{-0.3cm}
\begin{align}
L_B=\frac{w^2+h^2}{D} v_{\text{eff}},
\label{buffer_length}
\end{align}
where $\frac{w^2+h^2}{D}$ is the maximum time to travel over the radial direction. We define operator $\mathcal{G}[\cdot]$ to describe the concentration of product $S_k$, and according to \eqref{solution_cdr3}, $\mathcal{G}[\cdot]$ can be expressed as \begin{equation}
\begin{aligned}
\mathcal{G}[C_{S_{i_0}}(t),C_{S_{j_0}}(t),n]\triangleq \varphi [{(n-1)C_{{S_{i_0}}}(t)}/{n},{C_{{S_{j_0}}}(t)}/{n}]*H_n(nL_B+L_C,t)*H_n(L_R,t).\label{operator_G}
\end{aligned}
\end{equation}
For the same reaction, we define operator $\mathcal{R}[\cdot]$ to characterize the residual concentration of $S_i$. According to \eqref{solution_cdr1}, $\mathcal{R}[\cdot]$ can be expressed as
\begin{equation}
\begin{aligned}
\mathcal{R}[C_{S_{i_0}}(t),C_{S_{j_0}}(t),n]\triangleq &
\big[  {(n-1)C_{{S_{i_0}}}(t)}/{n}-\varphi [{(n-1)C_{{S_{i_0}}}(t)}/{n}, {C_{{S_{j_0}}}(t)}/{n}] \big]\\
& *H_n(nL_B+L_C,t)*H_n(L_R,t). \label{operator_R}
\end{aligned}
\end{equation}

For the amplifying reaction $S_i +Amp \to S_i+O$, operator $\mathcal{A}[\cdot]$ describes the concentration of product $O$, and can be expressed using \textbf{Lemma \ref{lemma_amplification}} as
\begin{equation}
\begin{aligned}
\mathcal{A}[C_{S_{i_0}}(t),C_{Amp_0}(t),n]\triangleq &\big[  
[{C_{Amp_0}(t)}/{n}*H_n(nL_B+L_C,t)]\\
&\cdot \mathbbm{1}_{ \{  [{(n-1)C_{S_{i_0}}(t)}/{n}*H_n(nL_B+L_C,t)]>0  \} }  
\big]*H_n(L_R,t). \label{operator_A}    
\end{aligned}
\end{equation}
As seen in the AND gate design in Fig. \ref{AND_gate_design}, a threshold reaction is cascaded with an amplifying reaction; thus, we define operator $\mathcal{F}[\cdot]$ as a combination of operators $\mathcal{R}[\cdot]$ and $\mathcal{A}[\cdot]$, which represents the concentration of product $O$ with $S_i +S_j \to S_k$ and $S_i +Amp \to S_i+O$ as
\begin{align}
\mathcal{F}[C_{S_i}(t),C_{S_j}(t),C_{Amp_0}(t),n]\triangleq \mathcal{A}\big[\mathcal{R}[C_{S_i}(t),C_{S_i}(t),n],C_{Amp_0}(t),n+1\big]. \label{operator_F}
\end{align}

\vspace{-0.7cm}
\subsection{AND Logic Gate Analysis}
We denote the concentrations of input species $I1$ and $I2$ as $C_{I1_0}(t)$ and $C_{I2_0}(t)$. Remind that we use non-zero concentration to represent HIGH state (bit-1), and zero concentration to represent LOW state (bit-0). Therefore, we assume that at any time $t$, $C_{I1_0}(t)$ and $C_{I2_0}(t)$ either are HIGH concentration $C_0$ or LOW concentration $0$.
Species $M$, $ThL$, and $Amp$ are injected continuously; thus, their initial concentrations follow $C_{M_0}(t)=C_{M_0}u(t)$, $C_{ThL_0}(t)=C_{ThL_0}u(t)$, and $C_{Amp_0}(t)=C_{Amp_0}u(t)$.
For simplicity, all reactants are injected using a same average velocity $v_{\text{eff}}$. 
 
\vspace{-0.3cm}
\begin{theorem} \label{theorem2} 
	The concentration distribution of the product species $O$ in our designed AND gate in Fig. \ref{AND_gate_design} can be derived as
	\vspace{-0.2cm}
	\begin{align} \label{AND_output}
	C_O(x_{5},t)=\mathcal{F}\bigl\{   \mathcal{T}[C_N(x_3,t),4],\mathcal{T}[C_{ThL_0}(t),1],\mathcal{T}[C_{Amp_0}(t),1],5\bigr\},
	\end{align}
	where 
	\begin{equation}
	\begin{aligned}
	\label{AND_x3}
	C_N(x_3,t)=&\frac{1}{2}\bigl\{ \mathcal{G}\big[
	\mathcal{T}[C_{I1_0}(t),1],\mathcal{T}[C_{M_0}(t),1],2\big]\\
	&+\mathcal{G}\big[ 
	\mathcal{T}[C_{I2_0}(t),1],\mathcal{T}[C_{M_0}(t),1],2\big] \bigr\}*H_2(L_{A2},t).
	\end{aligned}
	\end{equation}
	In \eqref{AND_output} and \eqref{AND_x3}, operators $\mathcal{T}[\cdot]$, $\mathcal{G}[\cdot]$, $\mathcal{F}[\cdot]$ are defined in Table \ref{operator}, $L_{A2}$ is the travelling distance of the laminar located at the centre channel from $x_2$ to $x_3$ in Fig. \ref{AND_gate_design}.
\end{theorem}
\vspace{-0.5cm}
\begin{proof} 
	As shown in Fig. \ref{AND_gate_design}, the species $N$ generated by two inputs merge with each other after reactions $I1+M\to N$ and $I2+M\to N$ at position $x=x_3$. With $L_{A2}$, the concentration of species $N$ at $x=x_3$ can be derived as \eqref{AND_x3}.
	Then, species $N$ travels over a convection-diffusion channel and enters the elementary block $\mathcal{F}[\cdot]$ consisting of reactions $N +ThL \to W$ and $N +Amp \to N+O$ to produce the gate output $O$. According to the definition of $\mathcal{F}[\cdot]$ in \eqref{operator_F}, the concentration of species $O$ at location $x=x_5$ can be derived as \eqref{AND_output}.
\end{proof}

\vspace{-0.2cm}
For the thresholding reaction $N+ThL\to W$ in Fig. \ref{AND_gate_design}, $C_{ThL_0}$ directly determines the gate function. We derive the constraint for $C_{ThL_0}$ in the following lemma.
\vspace{-0.3cm}
\begin{lemma} \label{lemma_ThL}
	To ensure that our designed AND gate exhibits desired behaviour, the concentration of species $ThL$ needs to satisfy
	\begin{align} \label{AND_constraint2}
	\lim_{t\to\infty} \frac{\varphi(C_0,C_{M_0})\mathcal{T}[u(t),1]*q(t)}{\mathcal{T}[u(t),1]*H_5(5L_{B}+L_C,t)}<C_{ThL_0}<2 \lim_{t\to\infty} \frac{\varphi(C_0,C_{M_0})\mathcal{T}[u(t),1]*q(t)}{\mathcal{T}[u(t),1]*H_5(5L_{B}+L_C,t)},
	\end{align}
	where $C_0$ is the HIGH concentration of input species $I1$ and $I2$, $C_{M_0}$ is the injected concentration of species $M$, $q(t)=H_2(L_{A2},t)*H_4(L_T,t)*H_5(5L_{B}+L_C,t)$, $H(x,t)$ is the transfer function of a convection-diffusion channel derived in \eqref{approx2}, and $\mathcal{T}[\cdot]$ is defined in \eqref{operator_T}.
\end{lemma}
\vspace{-0.4cm}
\begin{proof}
	Let $C_1$ and $C_{ThL}$ denote the steady-state concentrations of species $N$ and $ThL$ at location $x=x_4$, respectively. Fig. \ref{f_ThL} plots the concentration of species $N$ before and after reaction $N+ThL\to W$. 
	\begin{figure}
		\centering
		\includegraphics[width=4.6in]{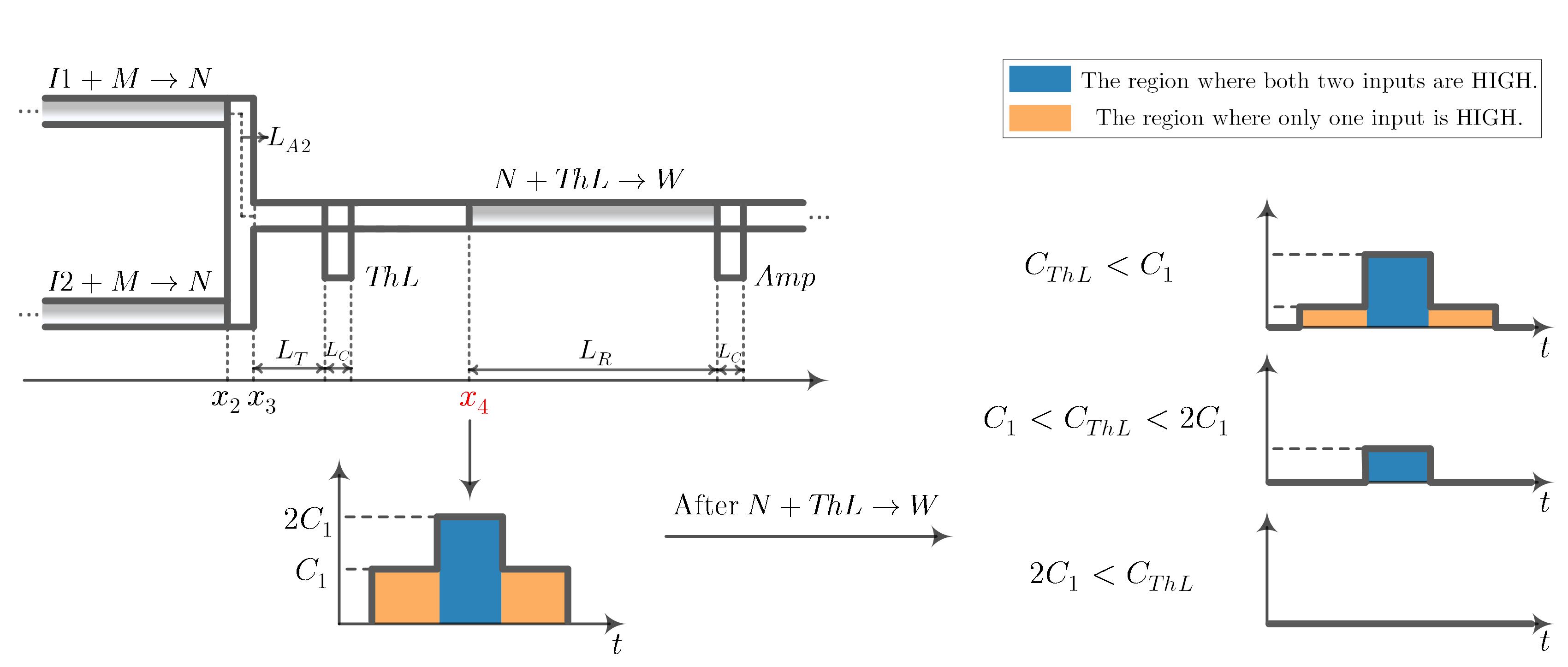}
		\caption{The concentration of species $N$ before and after the reaction $N+ThL\to W$ in the design AND gate.}
		\label{f_ThL}%
	\end{figure}	
	{When only one input is HIGH, the steady-state concentration $C_1$ can be expressed as}
	\vspace{-0.3cm}
	\begin{align} \label{AND_either_high}
	C_1=\lim_{t\to\infty} \frac{4}{5}\cdot \frac{1}{2} \cdot \mathcal{G}
	\bigl\{   	\mathcal{T}[C_0 u(t),1],\mathcal{T}[C_{M_0}(t),1],2\bigr\}*q(t).
	\end{align}
	When both inputs are HIGH, the steady-state concentration becomes $2C_1$. {For species $ThL$, its steady-state concentration $C_{ThL}$ at $x=x_4$ can be expressed as}
	\vspace{-0.2cm}
	\begin{align} \label{ThL_x4}
	C_{ThL}=\lim_{t\to\infty}\frac{1}{5}\mathcal{T}[C_{ThL_0} u(t),1]*H_5(5L_{B}+L_C,t).
	\end{align} 
	As shown in Fig. \ref{f_ThL}, the blue region represents that both two inputs are HIGH, and the yellow region represents that only one input is HIGH. The relationship between $C_1$ and $C_{ThL}$ has three cases:
	\begin{itemize}
		\item $C_{ThL}<C_1$: After reaction, the remaining concentration of species $N$ contains the region where one or both the inputs are HIGH.
		\item $C_1<C_{ThL}<2C_1$: After reaction, the remaining concentration of species $N$ only contains the region where both two inputs are HIGH.
		\item $2C_1<C_{ThL}$: After reaction, species $N$ is completely depleted.
	\end{itemize}
	Therefore, to capture the region where both the inputs are HIGH, the concentration of species $ThL$ needs to satisfy the condition $C_1<C_{ThL}<2C_1$. 
	Combined with \eqref{AND_either_high} and \eqref{ThL_x4}, we can derive \eqref{AND_constraint2}.
\end{proof}

\vspace{-1cm}
\section{Microfluidic QCSK Transmitter and Receiver}
\label{sec:QCSK}
In this section, we present the microfluidic designs to show how logic computations can process molecular concentration so as to achieve QCSK modulation and demodulation. Meanwhile, we also theoretically characterize the output concentration distributions of the proposed QCSK transmitter and receiver.

\vspace{-0.65cm}
\subsection{QCSK Transmitter}
\vspace{-0.2cm}
\subsubsection{QCSK Transmitter Design}
\begin{figure}
	\centering
	\includegraphics[width=2.3in]{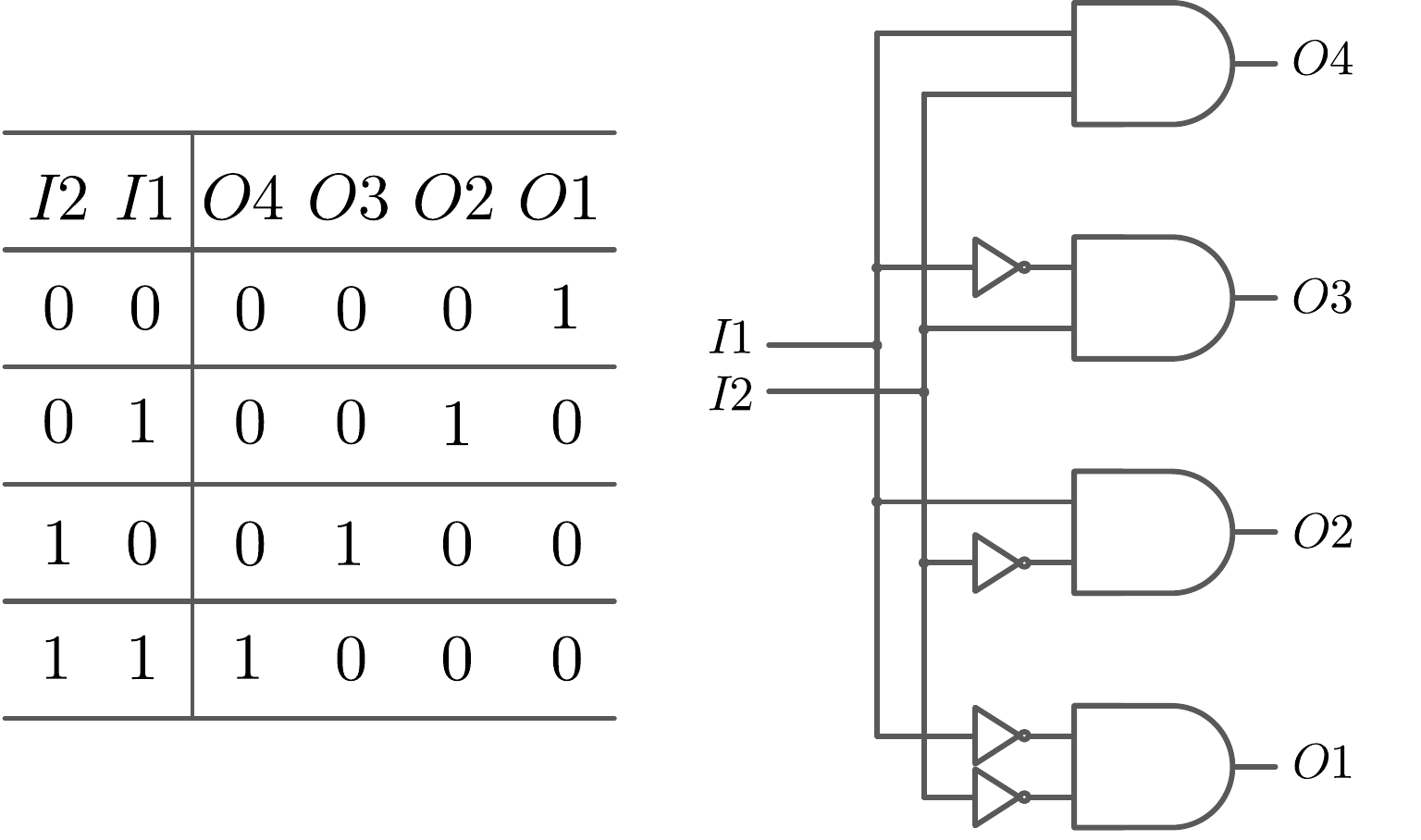}
	\caption{The truth table and implementation of an electric 2:4 decoder.}
	\label{f_decoder24}
\end{figure}

QCSK modulation represents two digital inputs as four concentration levels of an output signal, which is analogous to the Amplitude Shift Keying (ASK) modulation in wireless communication \cite{5962989}. A challenge of implementing a QCSK MC transmitter is how to control the output concentration via four different input combinations, \textit{i.e.}, ``00'', ``01'', ``10'', and ``11''. We solve this challenge by borrowing the idea of an electric 2:4 decoder. In electric field, a 2:4 decoder, which has 2 inputs and 4 outputs, selects exactly one of its outputs according to the input combination. Fig. \ref{f_decoder24} presents the truth table and an implementation for an electric 2:4 decoder, where four AND gates receive the HIGH or the LOW of input species $I1$ and $I2$. 

\begin{figure}
	\centering
	\includegraphics[width=5.5in]{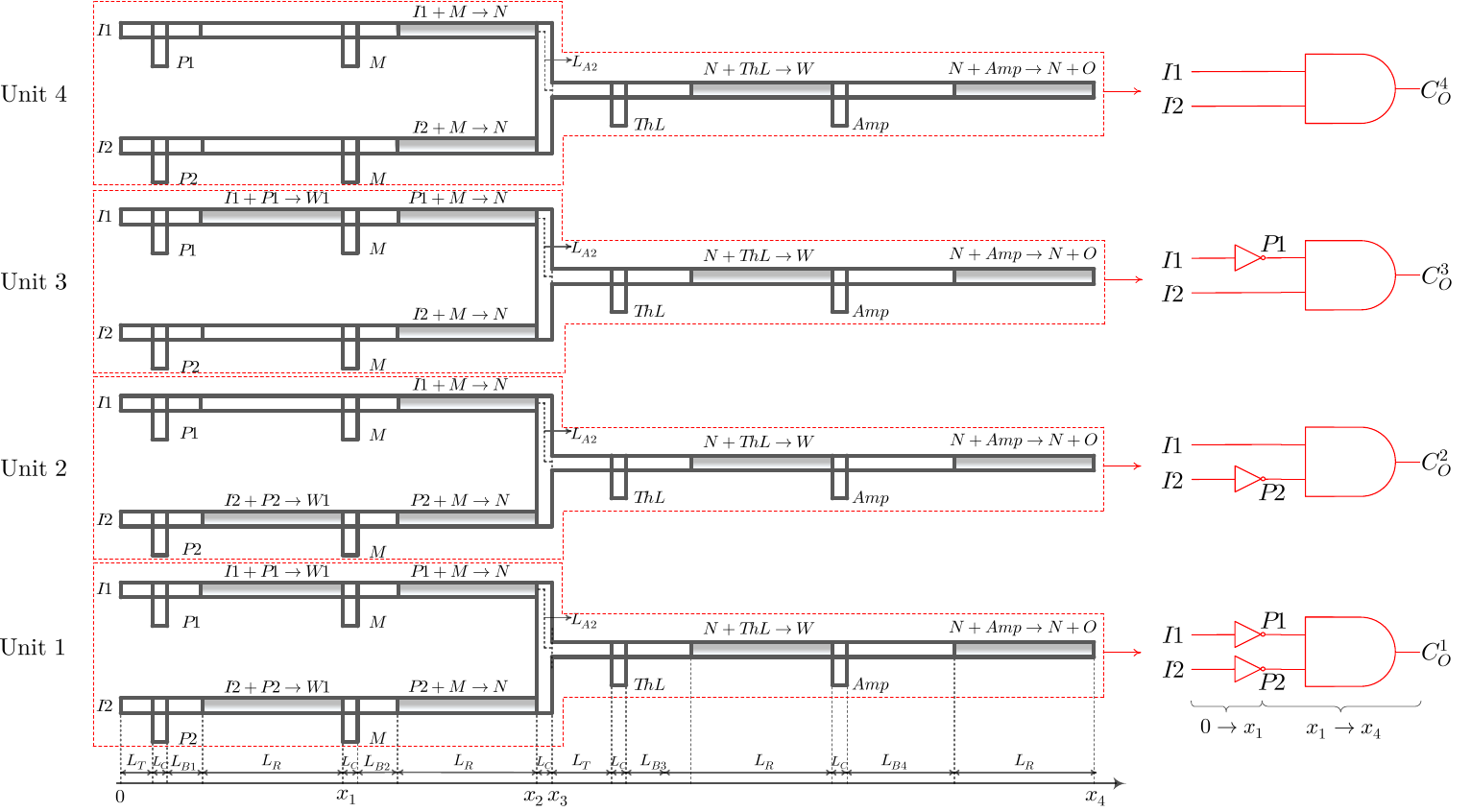}
	\caption{The chemical reactions-based microfluidic 2:4 decoder.}
	\label{f_decoder24_microfluidic}
\end{figure}

Inspired by the electric 2:4 decoder, we propose a chemical reactions-based microfluidic 2:4 decoder to realize QCSK modulation as Fig. \ref{f_decoder24_microfluidic}.
The proposed microfluidic device is made up of four microfluidic units corresponding to four different concentration outputs. For ease of reference, these four units are named as unit 4, unit 3, unit 2, and unit 1 from top to bottom. 
Analogous to the electric 2:4 decoder in Fig. \ref{f_decoder24}, the AND gate in each unit either takes $I1$ and $I2$ or their complementary species $P1$ and $P2$ as its inputs. It is noted that species $P1$ and $P2$ are supplied continuously with a HIGH state so that after reactions $I1+P1\to W1$ and $I2+P2\to W2$, the remaining concentrations of species $P1$ and $P2$, \textit{i.e.}, $C_{P1/P2}(x_1,t)$, can represent the complementary states of species $I1$ and $I2$, thus achieving the NOT gate. Unlike an electric 2:4 decoder that an identical voltage level is produced no matter which unit is selected, the proposed chemical 2:4 decoder will output different concentration levels. As each unit output $C_{O}^i(t)$ is influenced by $C_{Amp_0}^{i}$ in an amplifying reaction, the concentration variation of transmitted signals is represented via different concentrations of injected species $Amp$ as $C_{Amp_0}^{i}(t)=C_{Amp_0}^{i}u(t)$ ($1\le i \le 4$) for different units. Here, we set $C_{Amp_0}^{4}>C_{Amp_0}^{3}>C_{Amp_0}^{2}>C_{Amp_0}^{1}$ to ensure $\max{\{C_{O}^{4}(t)\}}>\max{\{C_{O}^{3}(t)\}}>\max{\{C_{O}^{2}(t)\}}>\max{\{C_{O}^{1}(t)\}}$.

\subsubsection{QCSK Transmitter Analysis}
\label{QCSK_TX_analysis}
The objective of the following analysis is to derive the transmitter output $C_{O}^i(t)$ of the design in Fig. \ref{f_decoder24_microfluidic}. We first derive the inputs of AND gates, \textit{i.e.}, the concentrations of $I1/I2$ and $P1/P2$ at location $x=x_1$. When input species $I1$ and $I2$ directly flow into an AND gate, their concentrations at location $x=x_1$ can be expressed as
\begin{align} \label{decoder24_x1_1}
	C_{I1/I2}(x_1,t)=\big[ \mathcal{T} [C_{{I1_0}/{I2_0}}(t),1]*H_2(L_C+L_{B1}+L_R,t)  \big]/2, 
\end{align}
where $C_{{I1_0}/{I2_0}}(t)$ is the concentration of input species $I1/I2$, operator $\mathcal{T}[\cdot]$ is defined in Table \ref{operator}. When the complementary species $P1$ and $P2$ flow into an AND gate, their concentrations at location $x=x_1$ can be expressed as
\vspace{-0.2cm}
\begin{align} \label{decoder24_x1_2}
	C_{P1/P2}(x_1,t)=\mathcal{R}
	\bigl\{\mathcal{T}[C_{{P1_0}/{P2_0}}(t),1],\mathcal{T}[C_{{I1_0}/{I2_0}}(t),1],2\bigr\},
\end{align}
where $C_{{P1_0}/{P2_0}}(t)$ is the concentration of species $P1/P2$, operator $\mathcal{R}[\cdot]$ is also defined in Table \ref{operator}.

With the derived AND gate inputs $C_{I1/I2}(x_1,t)$ in \eqref{decoder24_x1_1} and $C_{P1/P2}(x_1,t)$ in \eqref{decoder24_x1_2}, the transmitter output $C_{O}^i(t)$ can be expressed using \textbf{Theorem \ref{theorem2}} by interchanging the parameters 
\begin{itemize}
	\item in \eqref{AND_x3} via: $\mathcal{T}[C_{I1_0}(t),1] \to C_{I1/I2}(x_1,t)$ if an AND gate input is $I1/I2$, $\mathcal{T}[C_{I1_0}(t),1] \to C_{P1/P2}(x_1,t)$ if an AND gate input is $P1/P2$, $n=2 \to n=3$, $H_2(L_{A2},t) \to H_3(L_{A2},t)$;
	\item in \eqref{AND_output} via: $\mathcal{T}[C_{N}(x_3,t),4] \to \mathcal{T}[C_{N}(x_3,t),6]$, $C_{Amp_0}(t) \to C_{Amp_0}^i(t)$, $n=5 \to n=7$.
\end{itemize}

\vspace{-0.7cm}
\subsection{QCSK Receiver}
\vspace{-0.25cm}
\subsubsection{QCSK Receiver Design}
From the communication perspective, a corresponding microfluidic receiver is required to distinguish different concentration levels of $C_O^i(x_4,t)$ from different input combinations to achieve QCSK demodulation. In the following, we simplify $C_O^i(x_4,t)$ using $C_O^i(t)$ to represent a selected output of the proposed QCSK transmitter, and $C_O(t)$ to represent the general receiver input. In this paper, we consider the setup where the QCSK transmitter output $C_O^i(t)$ directly flows into the QCSK receiver; therefore, the receiver input 
$C_O(t)\in \big[0, \max{ \{C_O^4(t)\}} \big]$.
We also denote $C_Y^1(t)$ and $C_Y^2(t)$ as the final demodulated concentration signals, which correspond to the transmitter concentration inputs $C_{I1_0}(t)$ and $C_{I2_0}(t)$, respectively.

\begin{figure}
	\centering
	\includegraphics[width=5in]{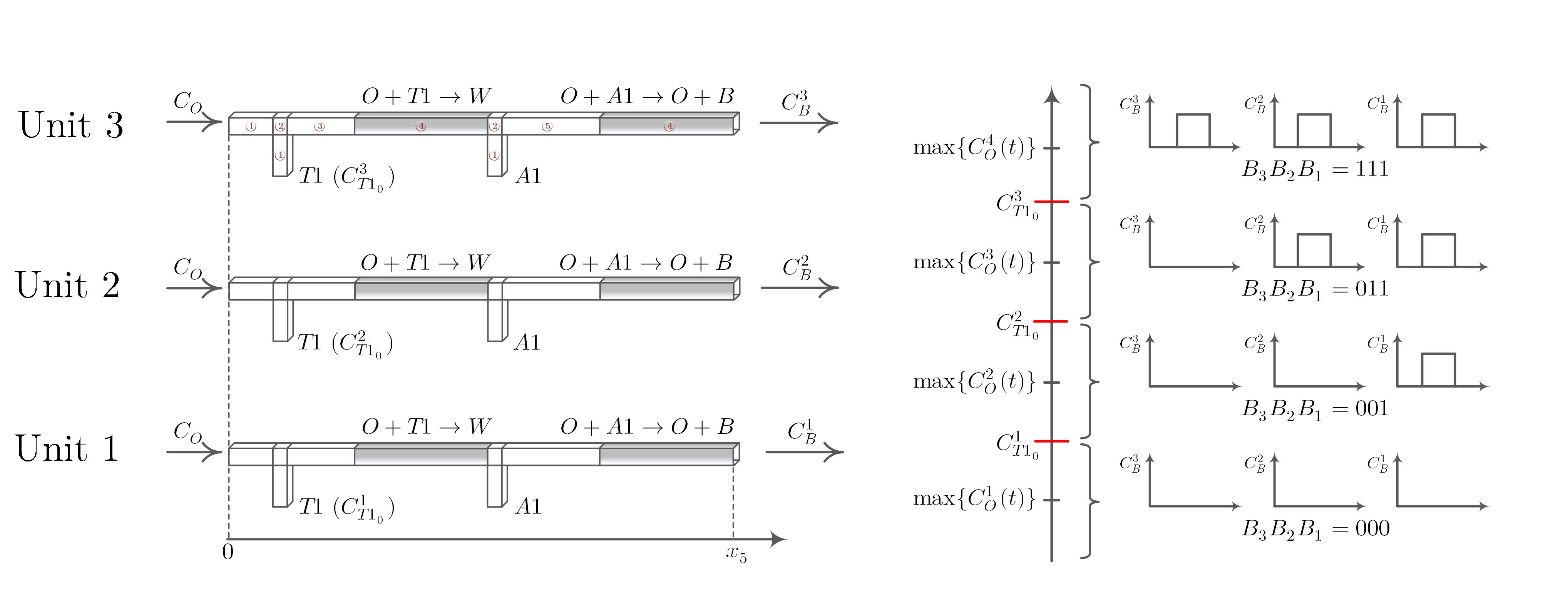}
	\caption{{Three detection units \cite{bi2019chemical} serve as a front-end processing modules. Each channel is labelled with a channel number to denote channel length as $L_{\text{number}}$. By setting $\max{\{C_O^{i}(t)\}}<C_{T1_0}^{i}<\max{\{C_O^{i+1}(t)\}}$, the front-end processing module can distinguish four concentration regions.
	}}
	\label{f_TCOM}
\end{figure}

To detect four concentration levels at the output of our proposed QCSK transmitter, we first design three detection microfluidic units in Fig. \ref{f_TCOM} to serve as a front-end processing module for the QCSK receiver, where each detection unit follows the receiver design in our initial work \cite{bi2019chemical}, with the capability of generating a rectangular output if the maximum concentration of a received signal exceeds a predefined threshold. 
As shown in Fig. \ref{f_TCOM}, the only difference among the three detection units is the injected concentration $C_{T1_0}^{i}(t)=C_{T1_0}^{i}u(t)$ $(1\le i \le3)$ of thresholding reactant $T1$. 
{{By setting $\max{\{C_O^{i}(t)\}}<C_{T1_0}^{i}<\max{\{C_O^{i+1}(t)\}}$, the concentration region of $C_O(t)$ can be identified for three-bit binary signals $B_3 B_2 B_1$ as shown in Fig. \ref{f_TCOM}. For instance, if $\max{\{C_O(t)\}}>C_{T1_0}^3$, all detection units will output a HIGH state with $B_3 B_2 B_1=111$. }}

\begin{figure*}[!tb]
	\centering
	\subfloat[Microfluidic channels to calculate $C_{Y}^2$\label{f_reactor1}]{\includegraphics[width=4.8in]{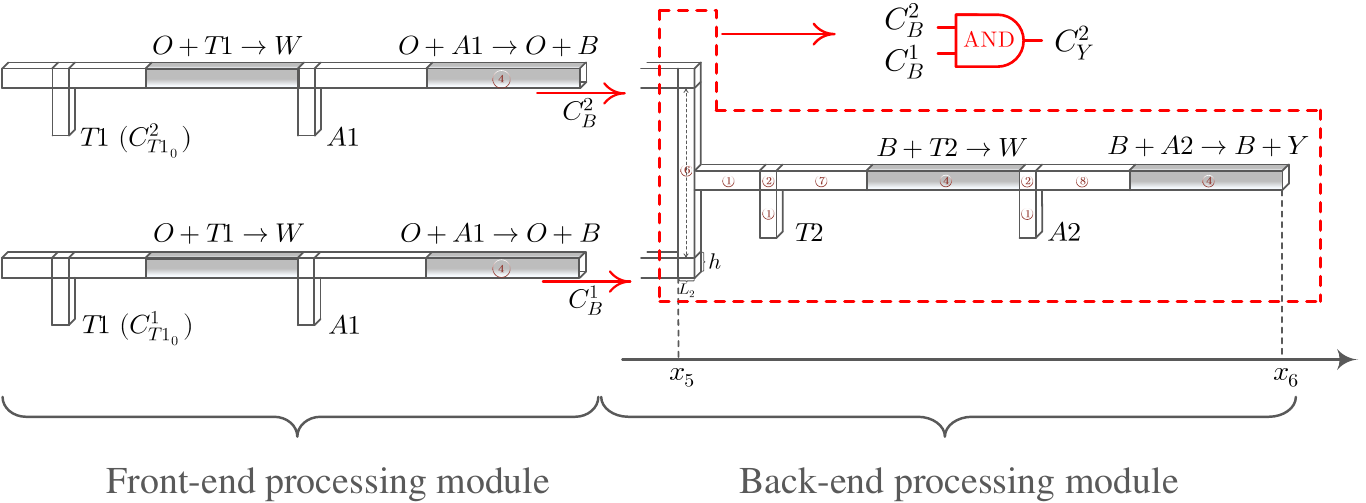}}
	\\
	\subfloat[Microfluidic channels to calculate $C_{Y}^1$\label{f_reactor2}]{\includegraphics[width=6.8in]{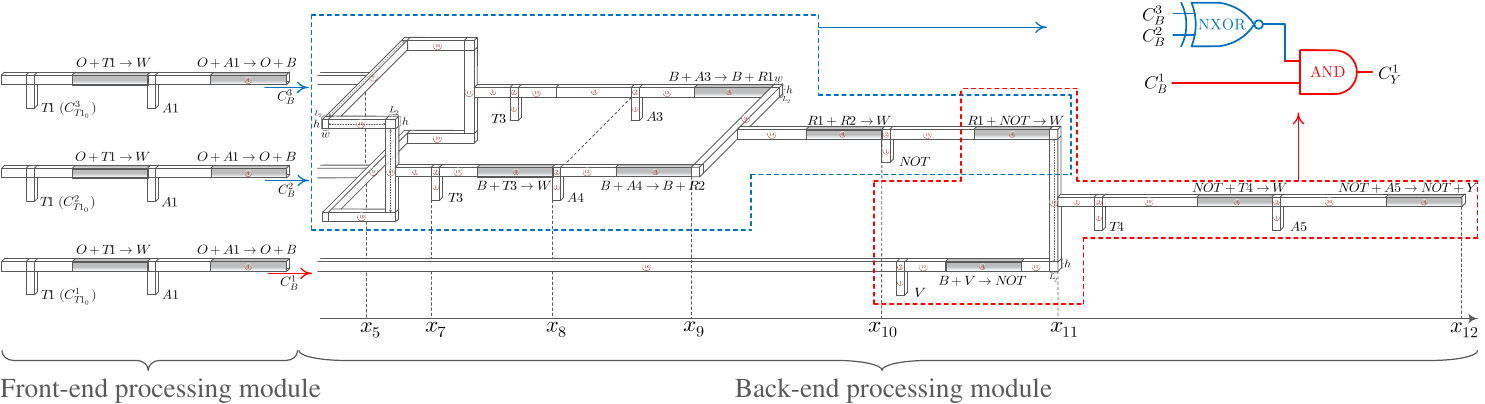}} 
	\caption{The microfluidic QCSK receiver design. Each channel is labelled with a channel number to denote channel length as $L_{\text{number}}$. The enlargement of (a) and (b) can see Appendix \ref{appendix_y2} and \ref{appendix_y1}, respectively.}
	\label{fig:QCSK_receiver}%
\end{figure*}

It is noted that the three detection units in Fig. \ref{f_TCOM} can only demodulate $C_O(t)$ to three concentration signals $C_B^3(t)$, $C_B^2(t)$, and  $C_B^1(t)$ instead of $C_{Y}^2(t)$ and $C_{Y}^1(t)$, which means extra signal processing units are required. Consider the outputs of front-end module $C_B(t)$ exhibit a rectangular concentration profile and its digital characteristic is ideal to perform logic computations \cite{Wang11}, this motivates us to design logic circuits to transform $C_B^i(t)$ to desirable output $C_{Y}^2(t)$ and $C_{Y}^1(t)$. 
To inspire the design for this signal transformation, we present the relationship between the binary signal $B_{i}$  ($1\le i \le 3$) and the binary signal $Y_{j}$  ($j=1,2$) in the truth table of Table \ref{rx_tt}.
\begin{table}[tb]
	\centering
	\caption{The relation between the receiver input $C_O(t)$, front-end module output binary signal $B$, and receiver output binary signal $Y$.}
	{\renewcommand{\arraystretch}{1}
		\scalebox{0.8}{\begin{tabular}{c ||c c c|c c}
				\hline
				\hline
				$\max{\{C_{O}(t)\}}$ & ${B}_3$  & ${B}_2$ & ${B}_1$  & ${Y}_2$ &${Y}_1$\\ \hline
				$[0, C_{{T1}_0}^1]$  & $0$ & $0$ & $0$ & $0$ & $0$ \\ 
				$[C_{{T1}_0}^1,C_{{T1}_0}^2]$  & $0$ & $0$ & $1$ & $0$ & $1$ \\ 
				$[C_{{T1}_0}^2,C_{{T1}_0}^3]$  & $0$ & $1$ & $1$ & $1$ & $0$ \\ 
				$[C_{{T1}_0}^3,\infty)$  & $1$ & $1$ & $1$ & $1$ & $1$ \\ 
				\hline
				\hline
		\end{tabular}}
	}
	\label{rx_tt}
\end{table} 
Based on Table \ref{rx_tt}, we express the Boolean equations \cite{harris2010digital} for $Y_2$ and $Y_1$ as 
\begin{equation} \label{rx_Y2}
\begin{split}
~~~~~~Y_2=\bar{B_3}{B_2}{B_1}+{B_3}{B_2}{B_1}
={B_2}{B_1},
\end{split}
\end{equation}
\begin{equation} \label{rx_Y1}
\begin{split}
\text{and}~~Y_1=\bar{B_3}\bar{B_2}{B_1}+{B_3}{B_2}{B_1}
={B_1}({B_3}\odot{B_2}),
\end{split}
\end{equation}
where $\bar{B_3}$ is the complementary form of $B_3$, ${B_2}{B_1}$ represents the AND operation of ${B_2}$ and ${B_1}$, and $\odot$ is the Exclusive NOR (XNOR) operation. 
Inspired by these Boolean relationships between species $B$ and receiver output species $Y$ in \eqref{rx_Y2} and \eqref{rx_Y1}, we connect the front-end module with an AND gate to compute $C_{Y}^2(t)$ as shown in Fig \ref{fig:QCSK_receiver}\subref{f_reactor1}, as well as a NXOR gate and an AND gate to calculate $C_{Y}^1(t)$ as shown in Fig. \ref{fig:QCSK_receiver}\subref{f_reactor2}.

\subsubsection{QCSK Receiver Analysis}
To theoretically characterize receiver outputs $C_Y^{2}(t)$ and $C_Y^{1}(t)$, {we denote $C_{{[\cdot]}_0}(t)$ as the concentration of any injected species $[\cdot]$, and $L_i$ as the length of the microfluidic channel with number $i$. Moreover, we assume that all types of molecules are injected with average velocity $v_\text{eff}$. In the following, we first derive the front-end processing output $C_B^{i}(t)$ in Fig. \ref{f_TCOM}, and then derive the QCSK receiver outputs $C_Y^{2}(t)$ and $C_Y^{1}(t)$ in Fig. \ref{fig:QCSK_receiver}. {In addition, the \textbf{location} and \textbf{channel number} are in bold in the following so that readers can easily follow our derivation.}

\bm{$C_B^{i}(t)$} \textbf{Derivation:} As shown in Fig. \ref{f_TCOM}, each detection unit in the front-end processing module consists of a thresholding reaction $O+T1\to W$ and an amplifying reaction $O+A1\to O+B$, and the output can be expressed using the operator $\mathcal{F}[\cdot]$ defined in Table \ref{operator} as
\vspace{-0.2cm}
\begin{align} \label{B_x5}
	C_{B}^i(x_5,t)=\mathcal{F}\bigl\{\mathcal{T}[C_O(t),1],\mathcal{T}[C_{T1_0}^i(t),1],\mathcal{T}[C_{A1_0}(t),1],2\bigr\},
\end{align}
where $C_O(t)$ is the receiver input concentration.

\bm{$C_Y^{2}(t)$} \textbf{Derivation:} As shown in in Fig. \ref{fig:QCSK_receiver}\subref{f_reactor1}, $C_{B}^2(t)$ and $C_{B}^1(t)$ flow into an AND gate to produce $C_Y^{2}(t)$. At $\bm{x=x_6}$, $C_Y^2(t)$ can be derived as
	\vspace{-0.2cm}
	\begin{equation} 
	\begin{aligned}
	C_{Y}^2(x_6,t)=\mathcal{F} \bigl\{  &\mathcal{T}[\frac{1}{2}\sum_{j=1}^{2} C_B^j(x_5,t)*H_3(\frac{2L_2+L_6+h}{2},t),6] \\       
	&\mathcal{T}[C_{T2_0}(t),1],\mathcal{T}[C_{A2_0}(t),1],7    \bigr\},    \label{B_x6}		
	\end{aligned}
	\end{equation}
where $1/2$ represents the dilution of $C_B^{1/2}(x_5,t)$ by $C_B^{2/1}(x_5,t)$, $H_n(x,t)$ is given in \textbf{Theorem 1} with $n$ indicating that the average velocity is $nv_{\text{eff}}$, and the operator $\mathcal{T}[\cdot]$ is defined in Table \ref{operator}.
	
\bm{$C_Y^{1}(t)$} \textbf{Derivation:} As shown in Fig. \ref{fig:QCSK_receiver}\subref{f_reactor2}, an XNOR gate and an AND gate are linked to the front-end processing module to produce 
$C_Y^1(t)$. 
\begin{itemize}
	\item \textbf{XNOR Gate Analysis}: Relying on the fluid separation analysis in \textbf{Lemma \ref{lemma_bifurcation_velocity}}, at \bm{$x=x_5$}, 
	$C_B^j(x_5,t)$ $(j=2,3)$ is equally separated from channel \bm{$9$} to channels \bm{$10$} due to the symmetrical microfluidic design from \bm{$x_5$} to \bm{$x_9$} in Fig. \ref{fig:QCSK_receiver}\subref{f_reactor2}, resulting in a velocity reduction from $3v_{\text{eff}}$ in channel \bm{$4$} with $O+A1\to O+B$ to $1.5v_{\text{eff}}$ in channels \bm{$10$}. In channels \bm{$11$}, the confluence of $C_B^3(x_5,t)$ and $C_B^2(x_5,t)$ occurs, and then is diluted by species $T3$ injected at \bm{$x_7$}. Subsequently, the outer fluid performs reaction $B+T3\to W$ to capture the region where both $C_B^3(x_5,t)$ and $C_B^2(x_5,t)$ are HIGH {as the second case in Fig. \ref{f_ThL}}, while the inner fluid flows	forward without this reaction. At \bm{$x=x_9$}, the amplifying products $R1$ and $R2$ after reactions $B+A3\to B+R1$ and $B+A4\to B+R2$ can be expressed as
	\vspace{-0.2cm}
	\begin{equation}
	\begin{aligned} \label{Y1_x7}
	&C_{R1}(x_9,t)=\\\mathcal{A} \bigl\{   &
	\underbrace{	\mathcal{T}[\frac{1}{2}\sum_{j=2}^{3} C_B^j(x_5,t)*H_{1.5}(\frac{3L_2+L_9+2L_{10}+L_{11}+h+2w}{2},t),3]*\frac{3}{4}H_4(L_2+L_{12}+L_4,t)}_{C_B^{\text{Inner}}(x_8,t)},\\
	&\mathcal{T}[C_{A3_0}(t),1],5 \bigr\},
	\end{aligned}
	\end{equation}
	\vspace{-0.2cm}
	\begin{equation}
	\begin{aligned} \label{Y2_x7}
	\text{and}~~C_{R2}(x_9,t)=&\mathcal{F} \bigl\{ 
	\underbrace{\mathcal{T}[\frac{1}{2}\sum_{j=2}^{3} C_B^j(x_5,t)*H_{1.5}(\frac{3L_2+L_9+2L_{10}+L_{11}+h+2w}{2},t),3]}_{C_B^{\text{Outer}}(x_7,t)}, \\&         \mathcal{T}[C_{T3_0}(t),1],\mathcal{T}[C_{A4_0}(t),1],4\bigr\},
	\end{aligned}
	\end{equation}
	{where the superscript ``Inner" and ``Outer" represent the outer and inter fluids from \bm{$x_7$} to \bm{$x_9$}}, and $3/4$ in \eqref{Y1_x7} represents the dilution of species $B$ by species $T3$.

	After reaction $R1+R2\to W$, the remaining species $R1$ at \bm{$x=x_{10}$} will be HIGH when either $C_B^3(x_5,t)$ or $C_B^2(x_5,t)$ is HIGH, thus achieving an XOR gate. Relying on \eqref{solution_cdr1} in \textbf{Lemma \ref{lemma_thresholding}}, the remaining concentration of species $R1$ is derived as
	\vspace{-0.15cm}
	\begin{equation}  \label{Y1_x8}
	\begin{aligned}
	C_{R1}(x_{10},t)=&\frac{1}{2}\bigl\{   
	C_{R1}(x_9,t)-\varphi[C_{R1}(x_9,t),C_{R2}(x_9,t)]\bigr\}\\
	&*H_{5}(\frac{L_2+L_9+2w}{2},t)*H_{10}(L_{14},t)*H_{10}(L_4,t),
	\end{aligned}
	\end{equation}
	where $\varphi[\cdot,\cdot]$ is given in \eqref{f_max}.
	The cascaded reaction $R1+NOT\to W$ functions as a NOT gate similar to the reaction $I1+P1\to W$ in the QCSK transmitter in Fig. \ref{f_decoder24_microfluidic} in order to achieve the XNOR gate. At \bm{$x=x_{11}$}, the concentration of $NOT$ can be expressed using the operator $\mathcal{R}[\cdot]$ defined in Table \ref{operator} as
	\vspace{-0.15cm}
	\begin{align} \label{NOT23_x9}
	C_{NOT}^{2\&3}(x_{11},t)=\mathcal{R} \bigl\{ \mathcal{T}[C_{NOT_0}(t),1],C_{R1}(x_{10},t),11\bigr\}*H_{11}(\frac{2L_2+L_{18}+h}{2},t),
	\end{align}
	where the superscript $2\&3$ represents the species $NOT$ generated by $C_B^2(x_5,t)$ and $C_B^3(x_5,t)$. 
	\item \textbf{AND Gate Analysis}: The calculation of receiver output $C_Y^1(t)$ also needs the participation of $C_B^1(x_5,t)$. To perform the AND gate, the product species $B$ (indicated by the red arrow) should be converted to molecular type $NOT$ via $B+V\to NOT$. At \bm{$x=x_{11}$}, the concentration of species $NOT$ generated by $C_B^1(x_5,t)$ can be expressed using operator $\mathcal{G}[\cdot]$ defined in Table \ref{operator} as 
	\vspace{-0.3cm}
	\begin{align} \label{NOT1_x9}
	C_{NOT}^1(x_{11},t)=\mathcal{G} \bigl\{   C_B^1(x_5,t)*H_3(L_{16},t),\mathcal{T}[C_{V_0}(t),1],4\bigr\}*H_4(\frac{2L_2+2L_{17}+L_{18}+h}{2},t).
	\end{align}
	We highlight that $C_{NOT}^1(x_{11},t)$ and $C_{NOT}^{2\&3}(x_{11},t)$ must be well synchronized. This means that $C_{NOT}^1(x_{11},t)$ and $C_{NOT}^{2\&3}(x_{11},t)$ should arrive at \bm{$x_{11}$} simultaneously, which can be achieved by ensuring the inputs $C_O(t)$ of three detection units have the same travelling time from the front-end module to position \bm{$x_{11}$} in Fig. \ref{fig:QCSK_receiver}\subref{f_reactor2}. Finally, we can derive the QCSK receiver output $C_{Y}^1(x_{12},t)$ as
	\vspace{-0.2cm}
	\begin{align} \label{Y_x10}
	C_{Y}^1(x_{12},t)=\mathcal{F} \bigl\{ \mathcal{T}[\frac{4}{15} C_{NOT}^1(x_{11},t)+\frac{11}{15}C_{NOT}^{2\&3}(x_{11},t),15],\mathcal{T}[C_{T4_0}(t),1],\mathcal{T}[C_{A5_0}(t),1],16 \bigr\},
	\end{align}
	where $4/15$ represents the dilution of $C_{NOT}^1(x_{11},t)$ by $C_{NOT}^{2\&3}(x_{11},t)$, while $11/15$ represents the dilution of $C_{NOT}^{2\&3}(x_{11},t)$ by $C_{NOT}^1(x_{11},t)$.
\end{itemize}

\vspace{-0.5cm}
\section{Performance Evaluation}
\label{sec:simulation}
\vspace{-0.2cm}
In this section, we implement our proposed microfluidic 
AND gate, QCSK transmitter, and QCSK receiver design in Fig. \ref{AND_gate_design}, Fig. \ref{f_decoder24_microfluidic}, and Fig. \ref{fig:QCSK_receiver} using COMSOL Multiphysics, which are then used to validate our corresponding theoretical analysis. The transfer function $H(x,t)$ given in \textbf{Theorem 1} is computed in Matlab using \textit{quadgk}. As \textit{quadgk} is only an approximation of $H(x,t)$, the computed results may fluctuate around their steady values. {If a computed value is slightly larger than steady value $0$, this can induce an instant change on the output value of the indicator function in \eqref{amp_approximation} from $0$ to $1$, which would further lead to a generation of output signals in undesired regions after an amplifying reaction. To avoid this phenomenon, we modify the statement of an indicator function $C_{S_{i_0}}(t)>0$ as $C_{S_{i_0}}(t)>\frac{1}{8}\max{\{C_{S_{i_0}}(t) \}}$.} By doing so, the width of a rectangular output is expected to be smaller than {that of the corresponding simulation result}. In COMSOL simulations, unless otherwise stated, we set $v_{\text{eff}}=0.1$cm/s, $D_\text{eff}=10^{-8}$m$^2$/s, $w=20\mu$m, $h=10\mu$m, $k=400$m$^3$/(mol$\cdot$s). Furthermore, we use “Ana.” and “Sim.” to abbreviate “Analytical” and “Simulation” in all figures.

\vspace{-0.6cm}
\subsection{AND Logic Gate}
\vspace{-0.2cm}
Fig. \ref{f:AND_gate} presents the COMSOL simulation results of the AND logic gate design depicted in Fig. \ref{AND_gate_design}. We set the parameters: $C_{I1_0}(t)=8[u(t-1)-u(t-3)]$, $C_{I2_0}(t)=8[u(t-2)-u(t-4)]$, $C_{M_0}(t)=8u(t)$, $C_{Amp_0}(t)=12u(t)$, $L_T=80\mu$m, $L_C=20\mu$m, $L_{R}=500\mu$m, $L_{A2}=120\mu$m. For the injected concentration of species $ThL$, we consider three cases: $C_{ThL_0}(t)=5u(t)$, $C_{ThL_0}(t)=10u(t)$, $C_{ThL_0}(t)=20u(t)$, with the aim to examine its impact on the gate behaviour.
\begin{figure*}[!tb]
	\centering	
	\subfloat[The concentration of species $N$ and $ThL$ at $x=x_4$.\label{ANDd1}]{\includegraphics[width=2.7in]{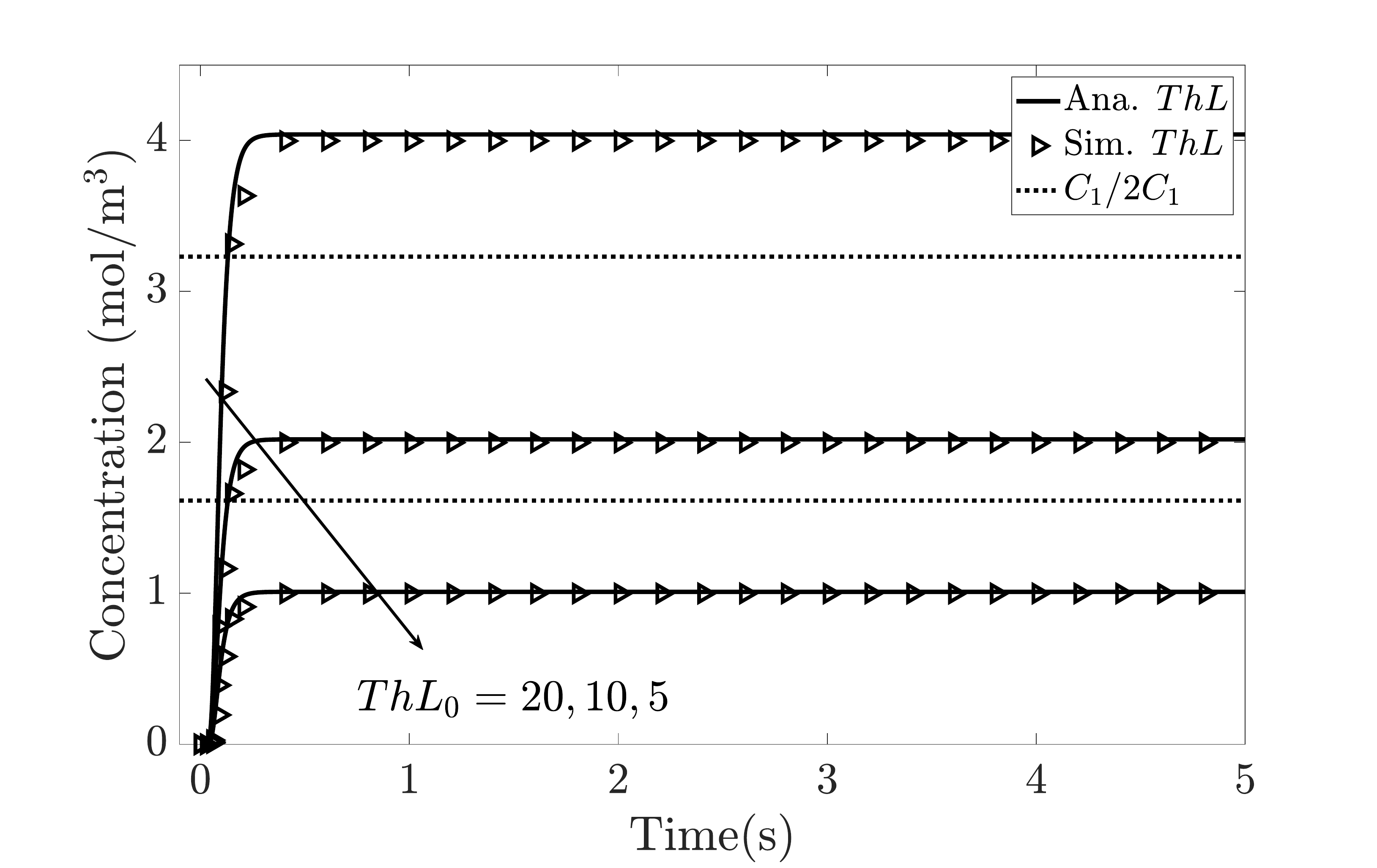}}%
	\qquad
	\subfloat[The normalized concentrations of input species $I1$, $I2$, and output species $O$.\label{ANDd3}]{\includegraphics[width=2.7in]{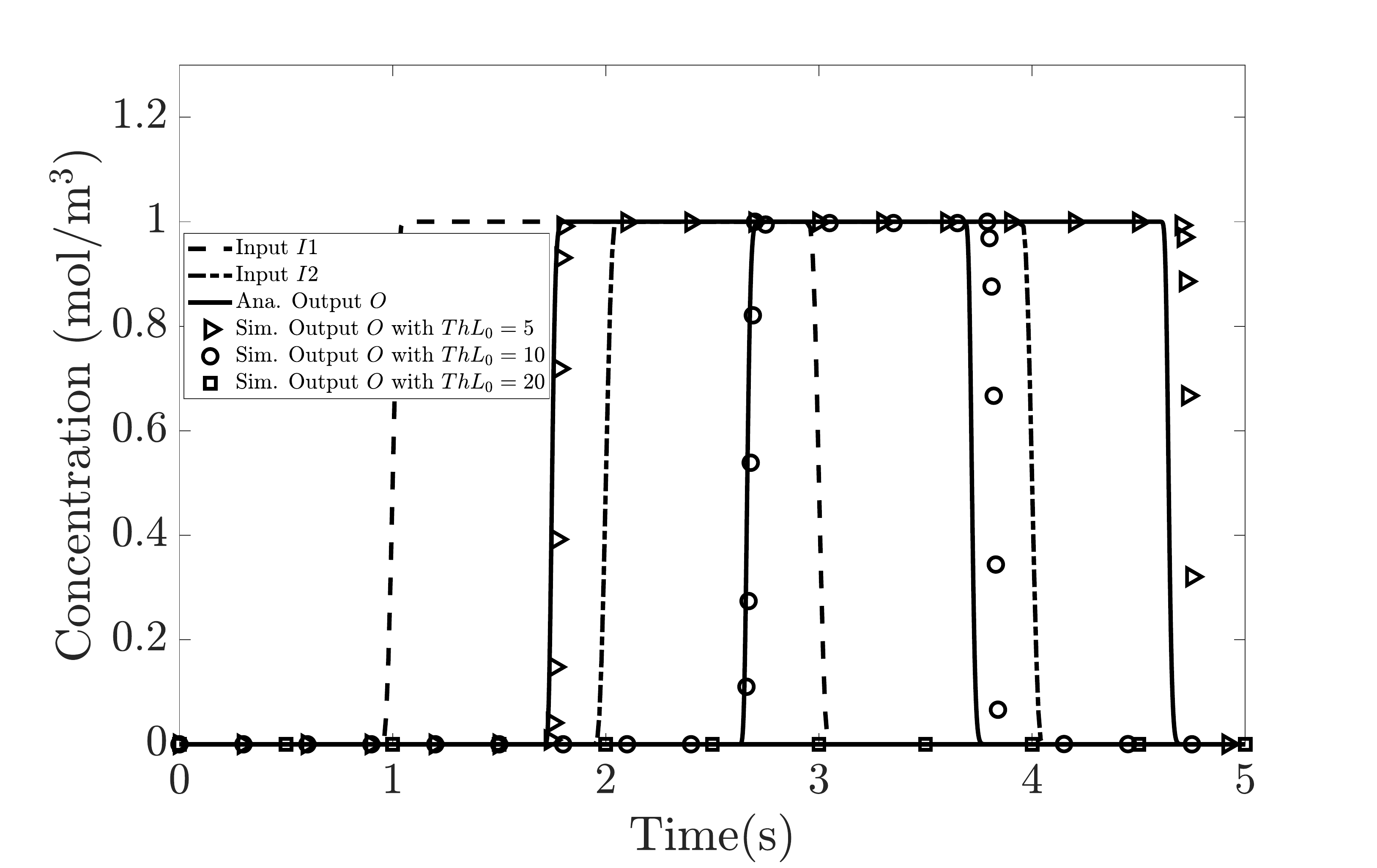}}
	\caption{The evaluation of an AND logic gate.}
	\label{f:AND_gate}%
\end{figure*}

\vspace{-0.7cm}
Fig. \ref{f:AND_gate}\subref{ANDd1} plots the concentrations of species $ThL$ before reaction $N+ThL\to W$ in Fig. \ref{AND_gate_design}. 
We observe that the simulated concentration points agree with the analytical concentration curves, thus demonstrating the correctness of our analysis of convection-diffusion in \textbf{Theorem 1} and convection-diffusion-reaction channels in \textbf{Lemma \ref{lemma_thresholding}}. 
For the three different injected concentrations, species $ThL$ is nearly diluted to one-fifth of its injected concentration due to that species $ThL$ enters the microfluidic device via the fifth inlet, which validates the concentration analysis for fluid convergence in {\textbf{Lemma \ref{lemma_combining_concentration}}}. Moreover, 
we also plot the concentration constraint $C_1$ in \eqref{AND_either_high} for species $ThL$ using black dash lines. 
For the curves with $C_{ThL_0}(t)=5u(t)$ or $20u(t)$, $C_{ThL_0}$ does not satisfy the concentration constraint in \textbf{Lemma \ref{lemma_ThL}}; as expected, the microfluidic device fails to achieve the AND function, which is demonstrated in Fig. \ref{f:AND_gate}\subref{ANDd3}. Fig. \ref{f:AND_gate}\subref{ANDd3} plots the normalized inputs and the final output product $O$ in \eqref{AND_output}. 
Only for $C_{ThL_0}(t)=10u(t)$, the width of species $O$ equals the width where both input species $I1$ and $I2$ are HIGH, demonstrating the desirable behaviour of an AND gate. Furthermore, due to the modification of the indicator function set, we can see the width of \eqref{AND_output} is smaller than that of the simulation results.

\vspace{-0.65cm}
\subsection{QCSK Transmitter}
\vspace{-0.2cm}
Fig. \ref{f:decoder} plots the outputs of the proposed microfluidic {QCSK transmitter design} in Fig. \ref{f_decoder24_microfluidic} 
and their analytical values $C_{O}^i(t)$ in Sec. \ref{QCSK_TX_analysis}. 
\begin{figure*}[!tb]
	\centering	
	\subfloat[Unit 4\label{decoder_branch4}]{\includegraphics[width=2.5in]{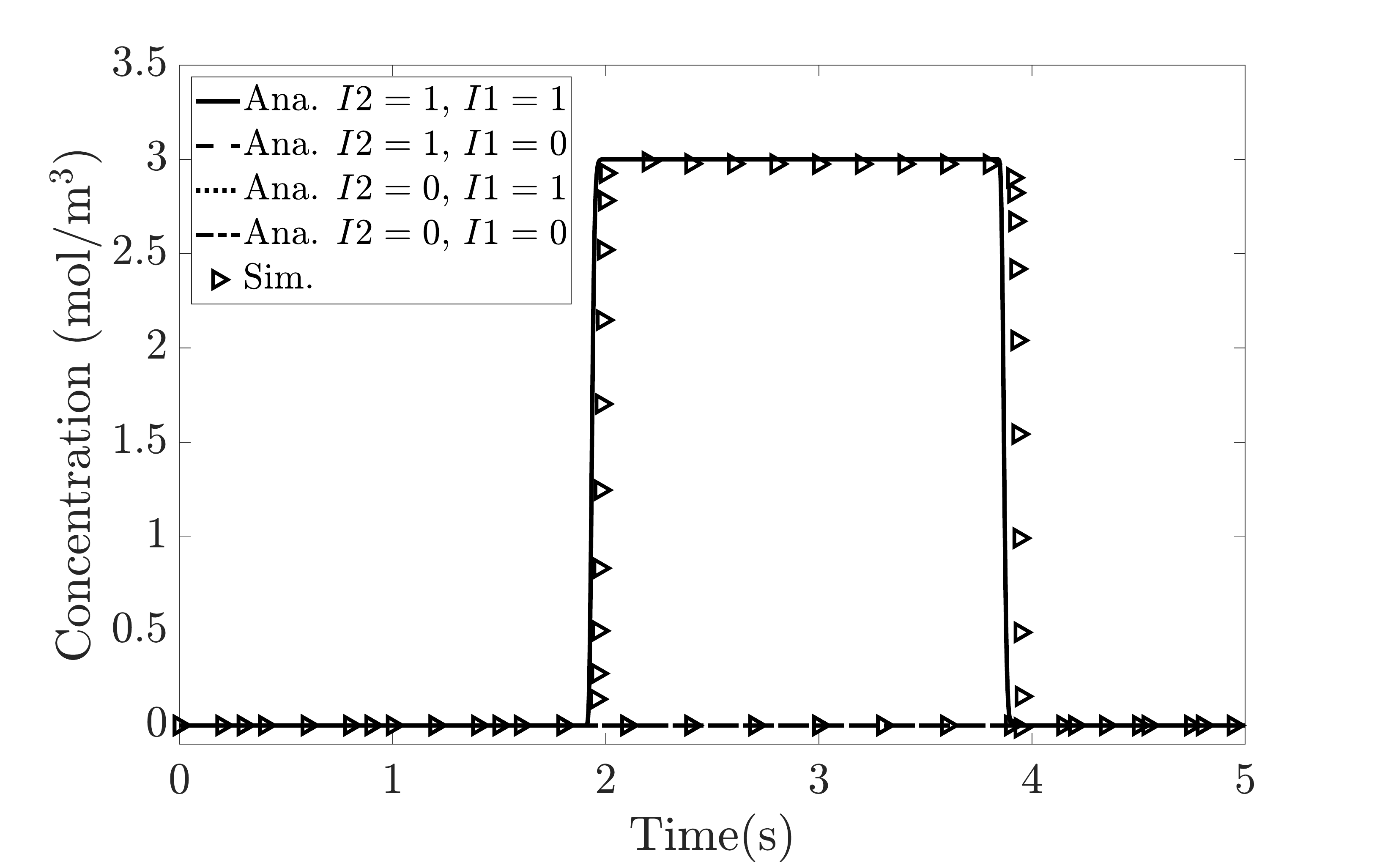}}%
	\qquad
	\subfloat[Unit 3\label{decoder_branch3}]{\includegraphics[width=2.5in]{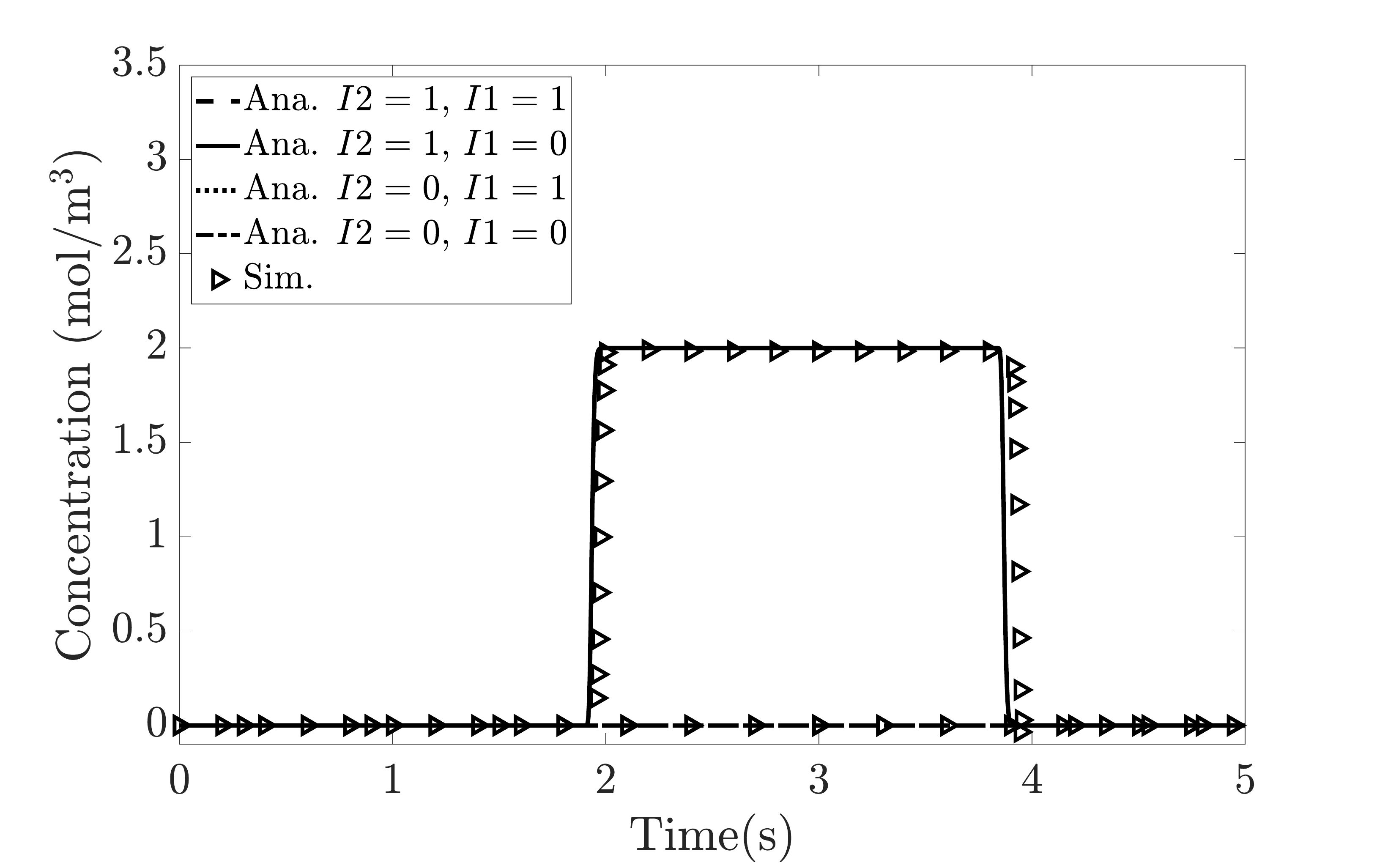}}
	\qquad 
	\subfloat[Unit 2\label{decoder_branch2}]{\includegraphics[width=2.5in]{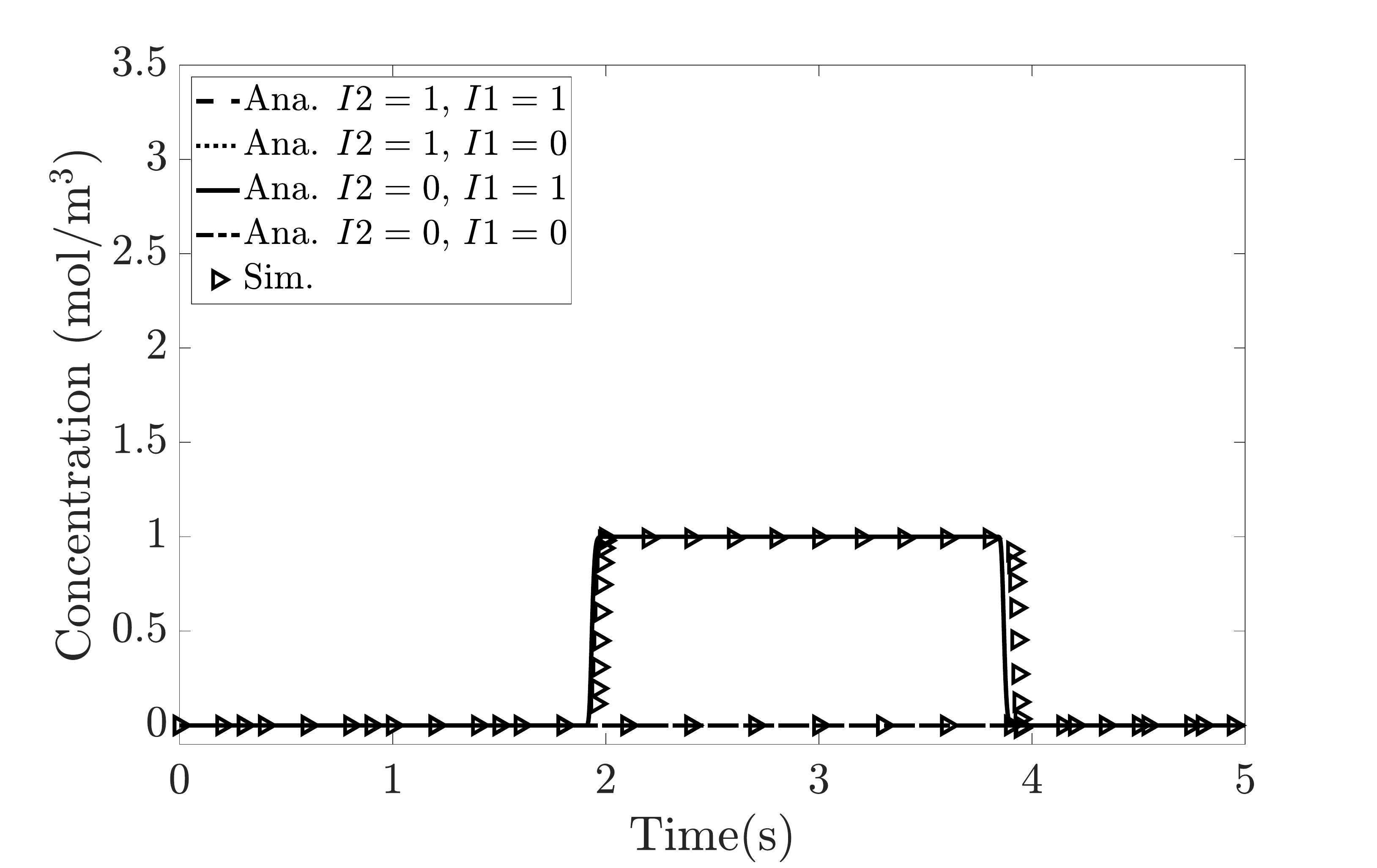}}
	\qquad 
	\subfloat[Unit 1\label{decoder_branch1}]{\includegraphics[width=2.5in]{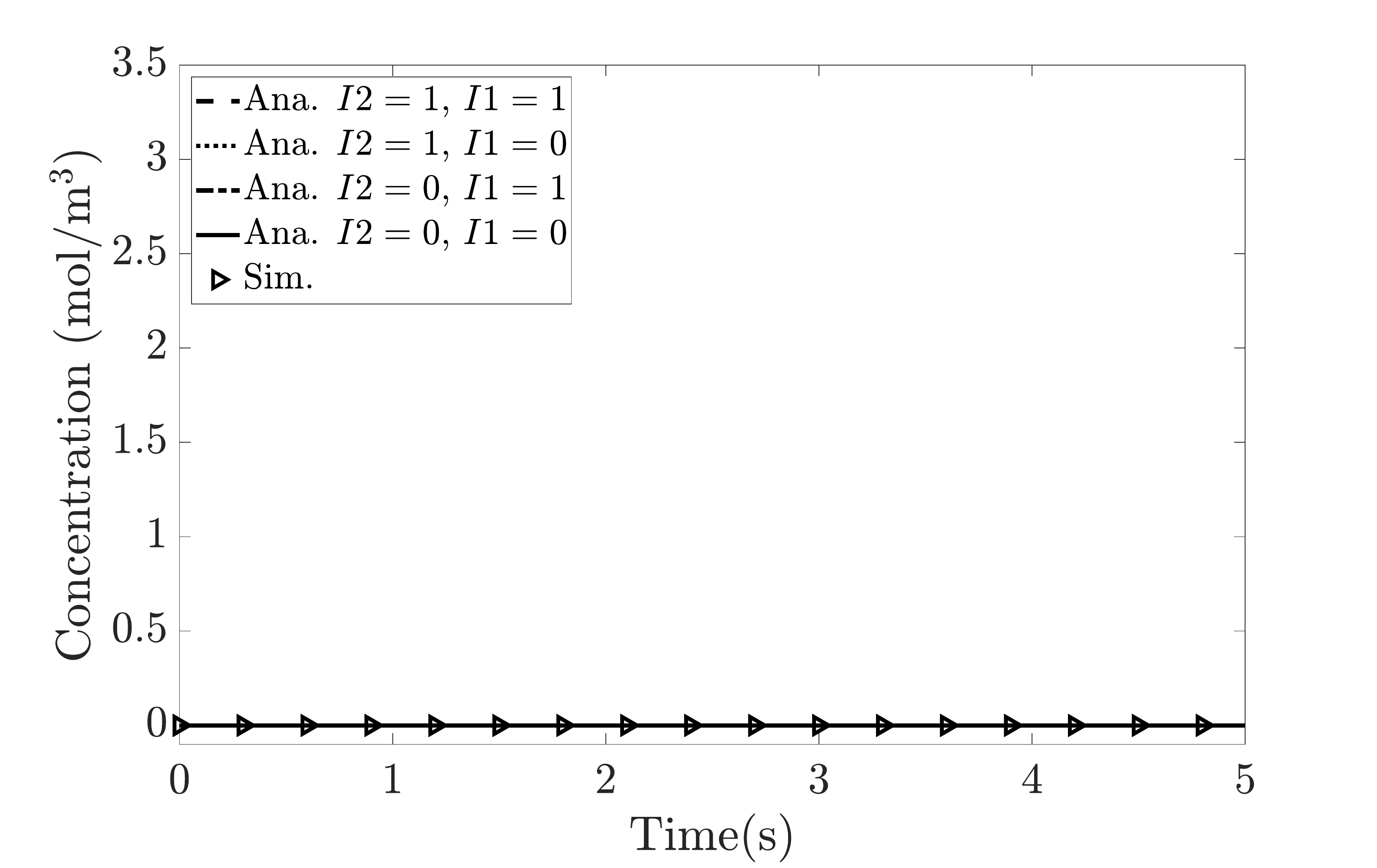}}
	\caption{The output concentrations of the proposed microfluidic QCSK transmitter.}
	\label{f:decoder}%
\end{figure*}
Species $I1$ and $I2$ are injected with either $12[u(t-1)-u(t-3)]$ representing bit $1$ or $0u(t)$ representing bit $0$. For other molecular types, their injected concentrations are set as: $C_{{P1_0}}(t)=C_{{P2_0}}(t)=12[u(t-1)-u(t-3)]$, $C_{M_0}(t)=12u(t)$, $C_{ThL_0}(t)=16u(t)$, $C_{Amp_0}^4(t)=24u(t)$, $C_{Amp_0}^3(t)=16u(t)$, $C_{Amp_0}^2(t)=8u(t)$, and $C_{Amp_0}^1(t)=0$. The buffer channels are configured with $L_{B1}=100\mu$m, $L_{B2}=150\mu$m, $L_{B3}=350\mu$m, and $L_{B4}=400\mu$m.

As shown in Fig. \ref{f:decoder}, for any input combination, only one unit outputs a HIGH signal except from the case where both $I1$ and $I2$ are LOW due to $C_{Amp_0}^1(t)=0$. Moreover, the analytical curves always capture the simulation points, which again demonstrates the effectiveness of our 
theoretical analysis $C_{O}^i(t)$ in Sec. \ref{QCSK_TX_analysis}. As species $Amp$ is supplied with different injected concentrations for each unit, we see that the selected unit reaches different concentration levels, proving that the proposed microfluidic QCSK transmitter successfully modulates input bits to the concentration level of output species $O$.

\vspace{-0.5cm}
\subsection{QCSK Receiver}
\vspace{-0.2cm}
To evaluate the proposed QCSK receiver design in Fig. \ref{fig:QCSK_receiver}, we consider four different rectangular concentration profiles as the receiver input $C_O(t)$, which is $C_O(t)=i[u(t-1)-u(t-3)]$ $(0\le i \le 3)$. Accordingly, to distinguish these four concentration levels, the concentration of species $T1$ for three units in Fig. \ref{f_TCOM} are set as:
{$C_{T1_0}^3(t)=2.5u(t)$, $C_{T1_0}^2(t)=1.5u(t)$, and 
$C_{T1_0}^1(t)=0.5u(t)$.}
Other parameters and the geometry are summarized in Table \ref{rx_parameter} and \ref{rx_geometry}.
\begin{table}[t]
	\centering
	\begin{minipage}{0.37\textwidth}
		\centering
		\caption{The parameters of the QCSK receiver.}
		{\renewcommand{\arraystretch}{1}
			\scalebox{0.5}{\begin{tabular}{c c||c c }
					\hline
					\hline
					Molecular Type & Concentration (mol/m$^3$)  & Molecular Type & Concentration (mol/m$^3$)  \\ \hline
					$A1$ & $9u(t)$     & $T2$ & $14u(t)$  \\
					$A2$ & $24u(t)$    & $T3$ & $7u(t)$  \\
					$A3$ & $20u(t)$    & $T4$ & $40u(t)$  \\
					$A4$ & $20u(t)$    & $NOT$ & $22u(t)$ \\
					$A5$ & $51u(t)$    & $V$ & $28u(t)$ \\
					\hline
					\hline
			\end{tabular}}
		}
		\label{rx_parameter}
	\end{minipage}
	\quad
	\begin{minipage}{0.6\textwidth}
		\centering
		\caption{The geometry of the QCSK receiver.}
		{\renewcommand{\arraystretch}{1}
			\scalebox{0.5}{\begin{tabular}{c c||c c ||c c||c c}
					\hline
					\hline
					Channel Number & Length ($\mu$m)  & Channel Number & Length ($\mu$m)  & Channel Number &Length ($\mu$m) & Channel Number &Length ($\mu$m)\\ \hline
					$1$ & $80$ & $6$ & $200$ & $11$ & $180$ & $16$ & $1911$ \\
					$2$ & $20$ & $7$ & $350$ & $12$ & $200$ & $17$ & $50$ \\
					$3$ & $100$ & $8$ & $400$ & $13$ & $250$ & $18$ & $300$ \\
					$4$ & $500$ & $9$ & $170$ & $14$ & $500$ & $19$ & $750$ \\
					$5$ & $150$ & $10$ & $180$ & $15$ & $550$ & $20$ & $800$ \\	
					\hline
					\hline
			\end{tabular}}
		}
		\label{rx_geometry}
	\end{minipage}
\end{table}

\begin{figure*}[!tb]
	\centering	
	\subfloat[{$C_O(t)=3[u(t-1)-u(t-3)]$}\label{rx_back11}]{\includegraphics[width=2.5in]{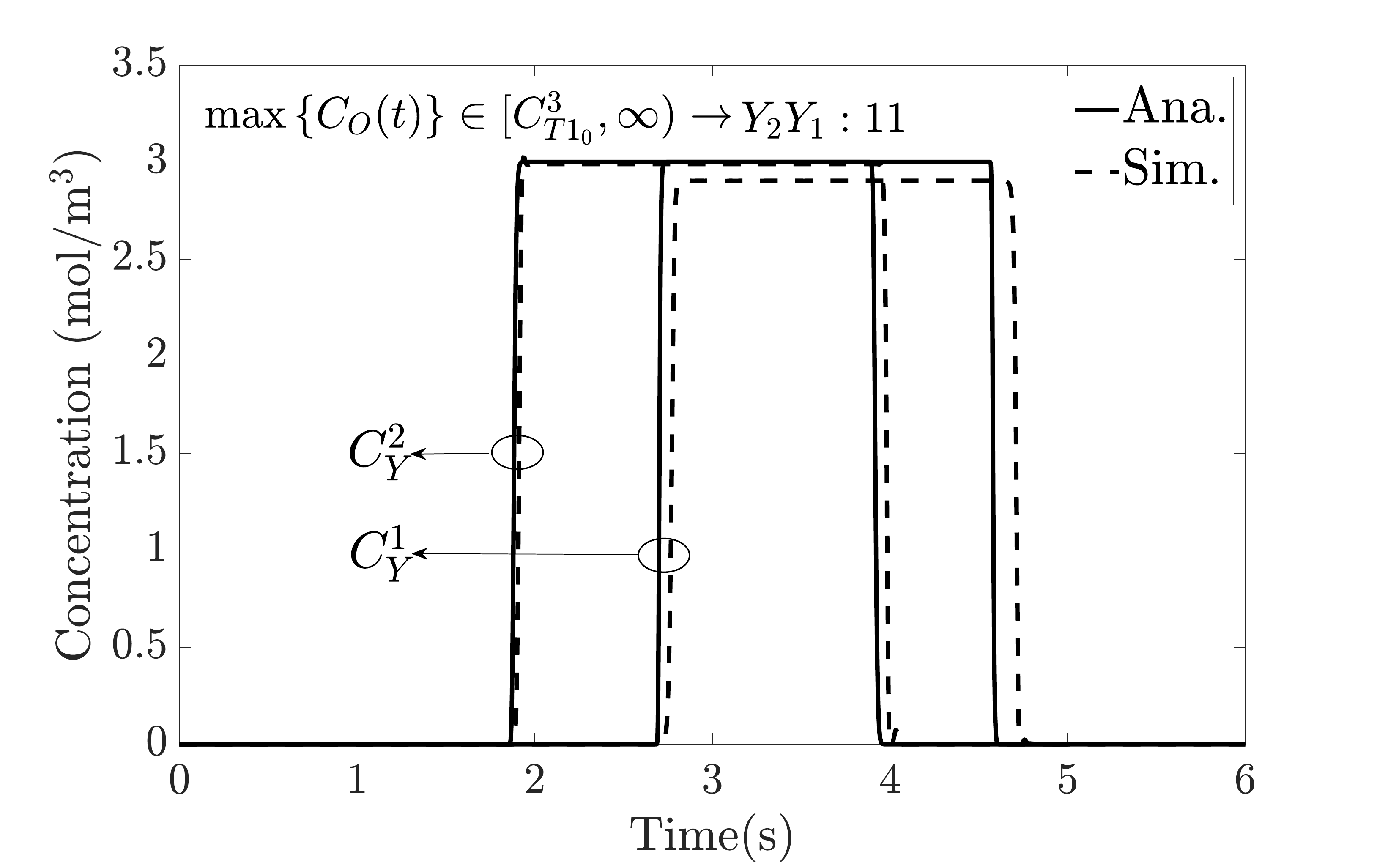}}%
	\qquad
	\subfloat[{$C_O(t)=2[u(t-1)-u(t-3)]$}\label{rx_back10}]{\includegraphics[width=2.5in]{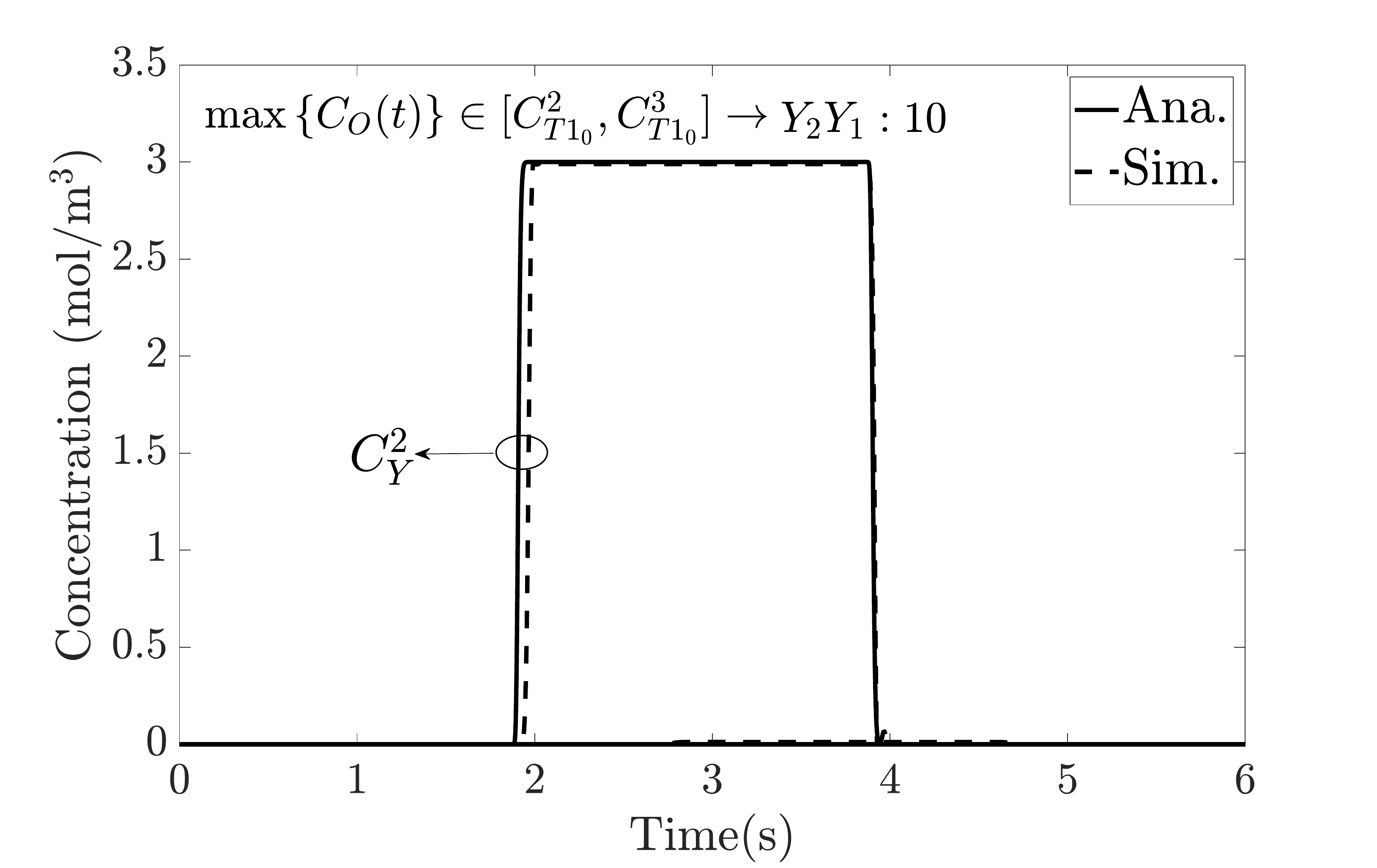}}
	\qquad 
	\subfloat[{$C_O(t)=u(t-1)-u(t-3)$}\label{rx_back01}]{\includegraphics[width=2.5in]{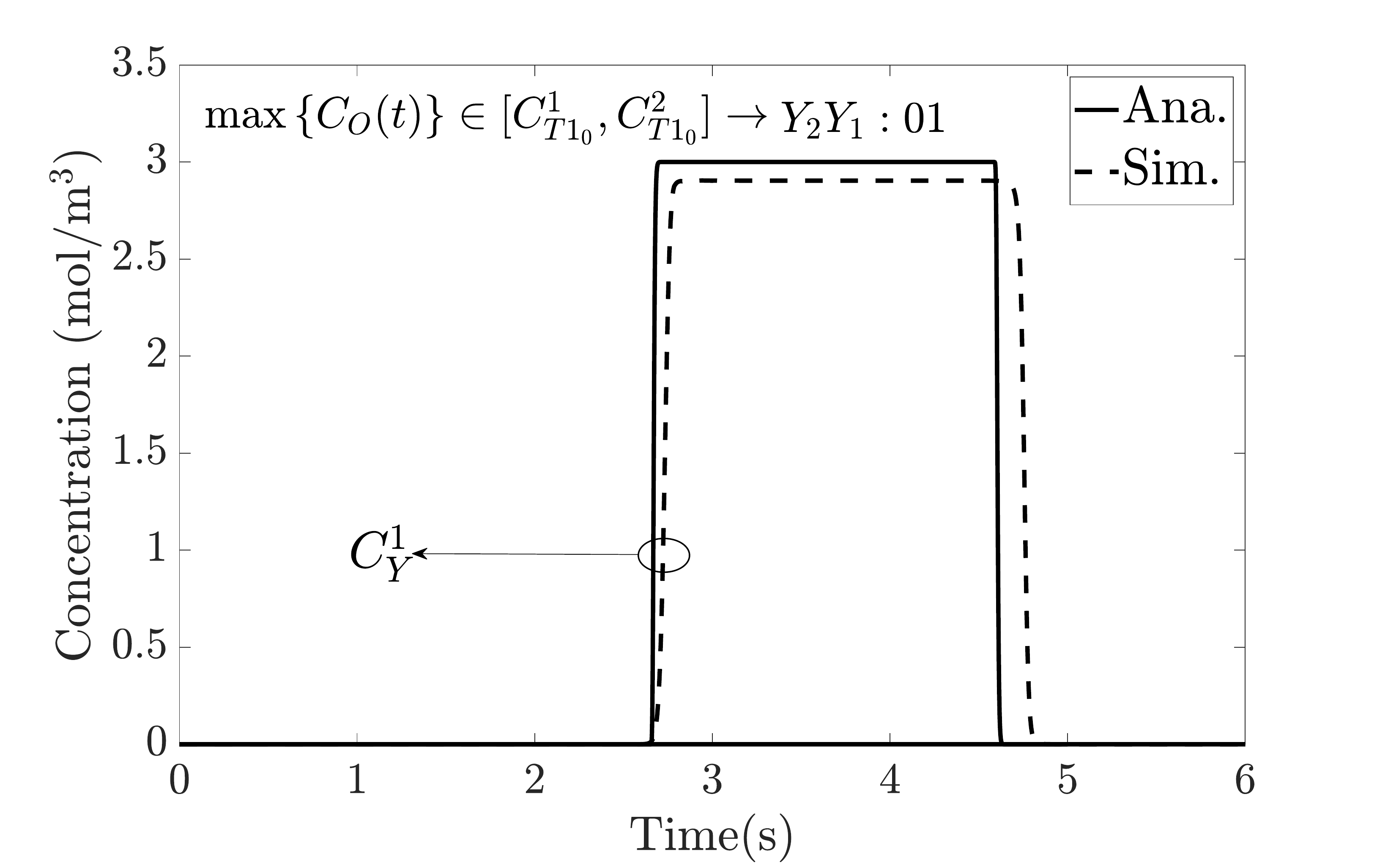}}
	\qquad 
	\subfloat[{$C_O(t)=0$}\label{rx_back00}]{\includegraphics[width=2.5in]{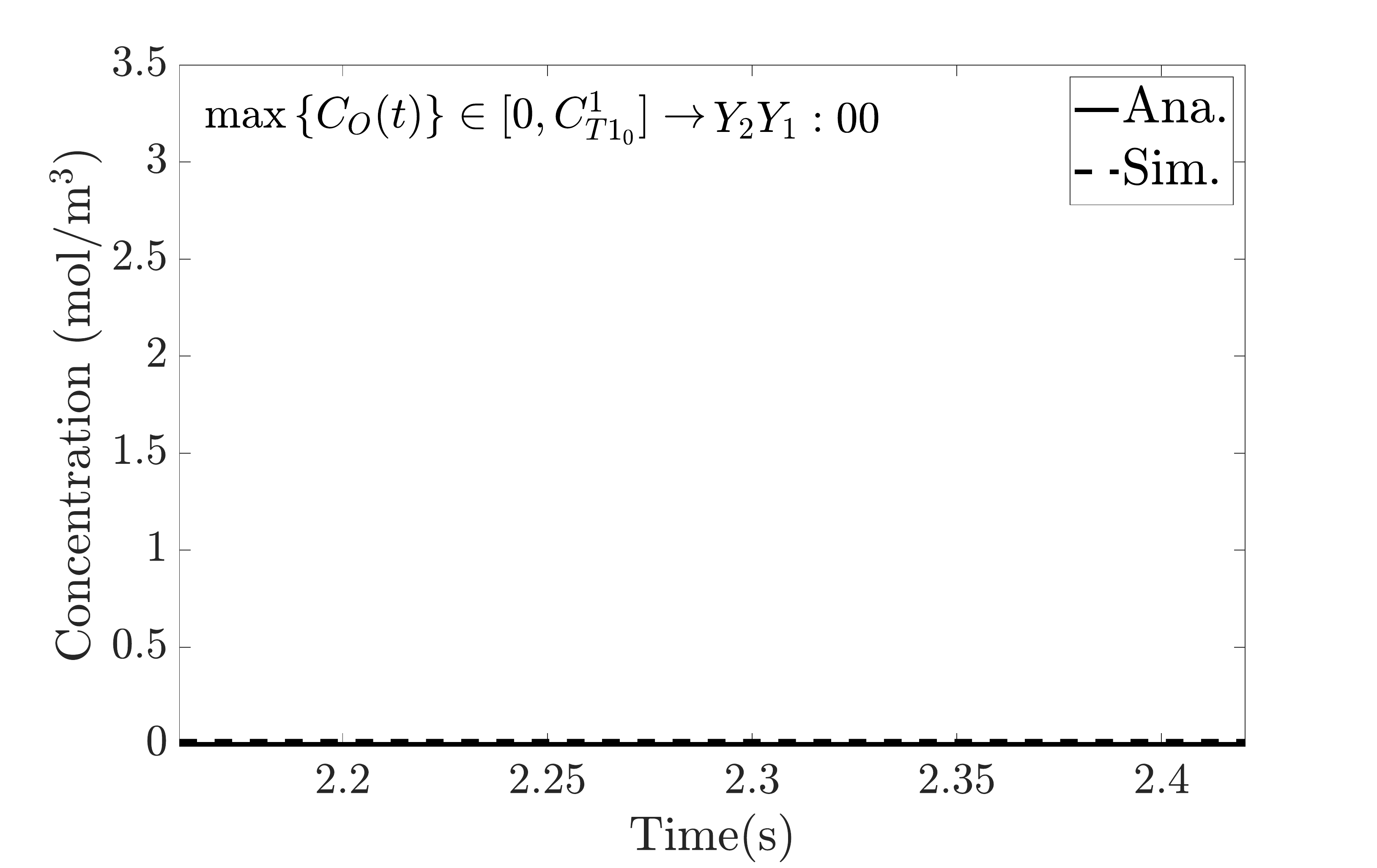}}
	\caption{The output concentrations of the proposed microfluidic QCSK receiver.}
	\label{rx_back}%
\end{figure*}
Fig. \ref{rx_back} plots the outputs of the proposed QCSK receiver design in Fig. \ref{fig:QCSK_receiver} and the corresponding analytical results of $C_Y^2(t)$ in \eqref{B_x6} and $C_Y^1(t)$ in \eqref{Y_x10}. First, we can see that although simulation curves are not in precise agreement with analytical curves, the close match can still confirm the correctness of the mathematical characterization of $C_Y^2(t)$ in \eqref{B_x6} and $C_Y^1(t)$ in \eqref{Y_x10}. Second, we observe the width difference between analytical and simulation curves for $C_Y^1$ is larger than that for $C_Y^2$. 
This is because the modification of the statement of an indicator function results in the  width difference in each amplifying reaction, the more amplifying reactions are utilized to compute $C_Y^1$ in Fig. \ref{fig:QCSK_receiver}\subref{f_reactor2}, the wider the width difference is.
Third, we see that the proposed receiver design can well demodulate the received signal $C_O(t)$ to two outputs $C_Y^2$ and $C_Y^1$. Recall that we use non-zero concentration to represent HIGH state (bit-1), and zero concentration to represent LOW state (bit-0). We also observe that the relationship between the maximum concentration of the receiver input $\max{\{C_O(t)\}}$, the concentration of species $T1$, and binary signals $Y_2$ and $Y_1$ is in consistent with the truth table of Table \ref{rx_tt}, which demonstrates the effectiveness of our proposed design.   

\vspace{-0.65cm}
\section{Conclusion}
\label{sec:con}
\vspace{-0.05cm}
In this paper, we considered the realization of quadruple concentration shift keying (QCSK) modulation and demodulation functionalities for molecular communication (MC) using chemical reactions-based microfluidic circuits. We first presented an AND gate design to demonstrate the logic computation capabilities of microfluidic circuits, and then showed how to utilize logic computations to achieve QCSK modulation and demodulation functions.
To theoretically characterize a microfluidic circuit, we established a general mathematical framework which is scalable with the increase of circuit scale and can be used to analyse other new and more complicated circuits. We derived the output concentration distributions of the AND gate, QCSK transmitter and receiver designs. 
Simulation results obtained from COMSOL Multiphysics showed all the proposed microfluidic circuits responded appropriately to input signals, and closely matched our derived analytical results. We believe that this paper not only provides a design principle and mathematical framework for microfluidic MC circuits, but also a foundation for harnessing simple microfluidic logic gates to produce diverse and complex signal processing functions.

%

\appendices
\vspace{-0.cm}
\section{Proof of Theorem 1}
\vspace{-0.3cm}
\label{app_cd}
To derive the transfer function $H(x,t)$, we formulate the following initial and boundary conditions for \eqref{cd}
\vspace{-0.6cm}
\begin{subequations}
	\begin{alignat}{2}
		C_{S_i}(0,t) = \delta(t),  \label{theo1_ini1}\\ 
		C_{S_i}(x,0) = 0,  ~x \ge  0, \label{theo1_ini2}\\ 
		\text{and} \;\; \frac{{\partial C_{S_i}(x ,t)}}{{\partial x}}|_{x=\infty}= 0,~  t \ge  0 \label{theo1_ini3}.
	\end{alignat}
\end{subequations}
\vspace{-0.2cm}{where $\delta(\cdot)$ is the Kronecker delta
function.} The Laplace Transform of \eqref{cd} with respect to $t$ is 
\begin{align}
{D_\text{eff}}\frac{{{\partial ^2}\widetilde{{C_{S_i}^{\text{}}}}(x,s)}}{{\partial {x^2}}} - {v_\text{eff}}\frac{{\partial \widetilde{{C_{S_i}^{\text{}}}}(x,s)}}{{\partial x}} - s\widetilde{{C_{S_i}^{\text{}}}}(x,s)=0. \label{second_order_ofe}
\end{align}
\vspace{-0.2cm}The general solution for this second order differential equation can be expressed as
\begin{align}
	\widetilde{{C_{S_i}^{\text{}}}}(x,s)=d_1 e^{\frac{{v_\text{eff}}+{\sqrt{ v_\text{eff}^2+4D_\text{eff}s }}}{2D_\text{eff}}x}+d_2 e^{\frac{{v_\text{eff}}-{\sqrt{ v_\text{eff}^2+4D_\text{eff}s }}}{2D_\text{eff}}x},
\end{align}
\vspace{-0.2cm}where $d_1$ and $d_2$ are two constants. To determine $d_1$ and $d_2$, we also apply Laplace Transforms to \eqref{theo1_ini1} and \eqref{theo1_ini3}, which are
\vspace{-0.2cm}
\begin{align}
	\widetilde{C_{S_i}}(0,s)=1, \label{re1}\\
	\frac{{\partial \widetilde{C_{S_i}}(x,s)}}{{\partial x}}|_{x=\infty}= 0. \label{re2}
\end{align}
Constrained by these two conditions, we arrive at the particular solution for \eqref{second_order_ofe} as
\begin{align}
\widetilde{{C_{S_i}^{\text{}}}}(x,s)=e^{\frac{{v_\text{eff}}-{\sqrt{ v_\text{eff}^2+4D_\text{eff}s }}}{2D_\text{eff}}x}, \label{parti}
\end{align}

\vspace{-0.2cm}
In order to obtain the transfer function, we need to calculate the inverse Laplace Transform of \eqref{parti}, \textit{i.e.}, $\mathcal{L}^{-1}\left\{     \widetilde{C_{S_i}}(x,s) \right\}$. However, $\mathcal{L}^{-1}\left\{     \widetilde{C_{S_i}}(x,s) \right\}$ is mathematically not solvable in close-form due to the complexity of \eqref{parti}. Here, we resort to the Gil-Pelaez theorem and consider $\mathcal{L}^{-1}\left\{     \widetilde{C_{S_i}}(x,s) \right\}$ as a probability density function whose characteristic function is $\widetilde{{C_{S_i}^{\text{}}}}(x,s)$. The cumulative distribution function (CDF) for $\mathcal{L}^{-1}\left\{     \widetilde{C_{S_i}}(x,s) \right\}$ can be expressed as
\vspace{-0.1cm}
\begin{align} \label{er15}
F(x,t)=\frac{1}{2}-\frac{1}{\pi}\int_{0}^{\infty} \frac{e^{-j\omega t}\overline{\widetilde{{C_{S_i}^{\text{}}}}(x,\omega)}-e^{j\omega t}{\widetilde{{C_{S_i}^{\text{}}}}(x,\omega)}}{2j\omega}\mathrm{d}w.
\end{align}
Take the derivative of $F(x,t)$ with respect to $t$,  we can arrive at \eqref{approx2}.

\vspace{-0.cm}
\section{The derivation of the Concentration of Product Species $S_k$ in \eqref{solution_cdr3}}
\vspace{-0.2cm}
To derive the concentration of product $S_k$, we combine \eqref{cdr1} and \eqref{cdr3} and denote $C(x,t)=C_{S_i}(x,t)+C_{S_k}(x,t)$, which yields
\begin{align}
\frac{\partial C(x,t)}{{\partial}t}=D_{\text{eff}}\frac{\partial^2 C(x,t)}{{\partial}x^2}-v_{\text{eff}}\frac{\partial C(x,t)}{{\partial}x}. \label{cdr4}
\end{align}
The sum concentration has the following initial and boundary conditions
\vspace{-0.3cm}
\begin{subequations}
	\begin{alignat}{2}
	C(0,t) = C_{{S_{i_0}}}(0,t),  \label{cdr4_ini1}\\ 
	C(x,0) = 0,  ~x \ge  0, \label{cdr4_ini2}\\ 
	\text{and} \;\; \frac{{\partial C(x ,t)}}{{\partial x}}|_{x=\infty}= 0,~  t \ge  0 \label{cdr4ini3}.
	\end{alignat}
\end{subequations}
{As these conditions are the same as \eqref{theo1_ini1}-\eqref{theo1_ini3}, we can write }
\vspace{-0.2cm}
\begin{align}
C(x,t)=C_{{S_{i_0}}}(0,t)*H(x,t).
\end{align} 
\vspace{-0.2cm}Combined with \eqref{surplus_i} and \eqref{solution_cdr1}, the concentration of product $S_k$ is
\begin{equation}
	\begin{aligned}
	C_{S_k}(x,t)&= C(x,t)-C_{S_i}(x,t)\\
	&= \varphi[C_{S_{i_0}}(t), C_{S_{j_0}}(t) ]*H(x,t).
	\end{aligned}
\end{equation}

\vspace{5cm}
\section{The Enlargement of the back-end processing module in Fig. 9({\upshape a})}
\label{appendix_y2}
\begin{figure}[H]
	\centering
	{\includegraphics[angle=270,width=2.8in]{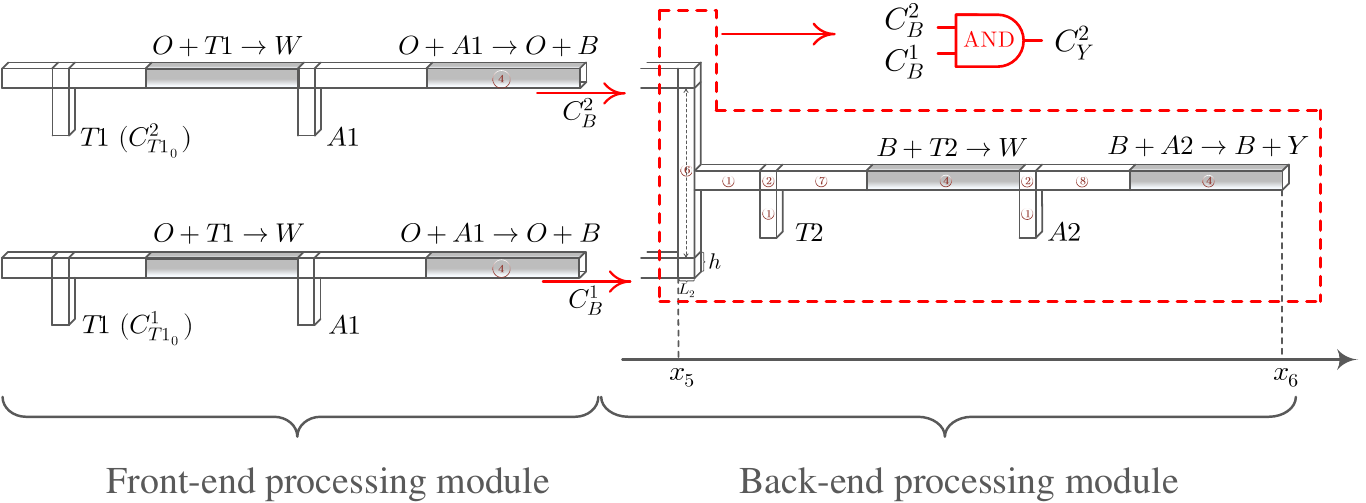}}
	\caption{Microfluidic channels for the QCSK output $C_{Y}^2$.}
\end{figure} 

\section{The Enlargement of the back-end processing module in Fig. 9({\upshape b})}
\label{appendix_y1}
\begin{figure}[H]
	\centering
	{\includegraphics[angle=270,width=2.1in]{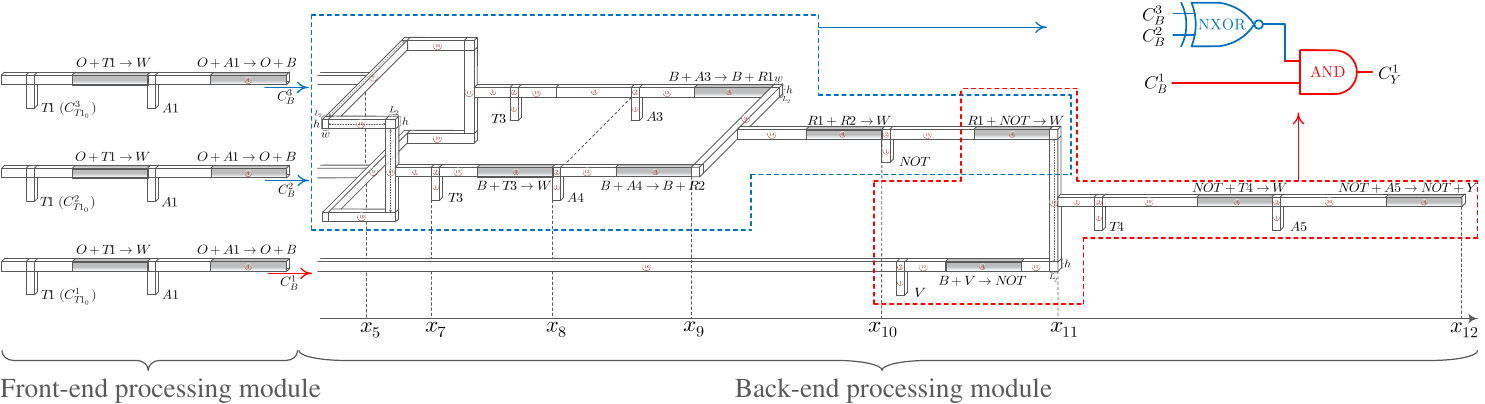}}
	\caption{Microfluidic channels for the QCSK output $C_{Y}^1$.}
\end{figure} 

\ifCLASSOPTIONcaptionsoff
  \newpage
\fi



\vspace{-0.6cm}


\begin{thebibliography}{10}
\baselineskip 12pt
\providecommand{\url}[1]{#1}
\csname url@samestyle\endcsname
\providecommand{\newblock}{\relax}
\providecommand{\bibinfo}[2]{#2}
\providecommand{\BIBentrySTDinterwordspacing}{\spaceskip=0pt\relax}
\providecommand{\BIBentryALTinterwordstretchfactor}{4}
\providecommand{\BIBentryALTinterwordspacing}{\spaceskip=\fontdimen2\font plus
\BIBentryALTinterwordstretchfactor\fontdimen3\font minus
  \fontdimen4\font\relax}
\providecommand{\BIBforeignlanguage}[2]{{%
\expandafter\ifx\csname l@#1\endcsname\relax
\typeout{** WARNING: IEEEtran.bst: No hyphenation pattern has been}%
\typeout{** loaded for the language `#1'. Using the pattern for}%
\typeout{** the default language instead.}%
\else
\language=\csname l@#1\endcsname
\fi
#2}}
\providecommand{\BIBdecl}{\relax}
\BIBdecl

\bibitem{8726417}
M.~{Kuscu}, E.~{Dinc}, B.~A. {Bilgin}, H.~{Ramezani}, and O.~B. {Akan},
  ``Transmitter and {R}eceiver {A}rchitectures for {M}olecular
  {C}ommunications: {A} {S}urvey on {P}hysical {D}esign {W}ith {M}odulation,
  {C}oding, and {D}etection {T}echniques,'' \emph{Proc. IEEE}, vol. 107, no.~7,
  pp. 1302--1341, July 2019.

\bibitem{8742793}
V.~{Jamali}, A.~{Ahmadzadeh}, W.~{Wicke}, A.~{Noel}, and R.~{Schober},
  ``Channel {M}odeling for {D}iffusive {M}olecular {C}ommunication—{A}
  {T}utorial {R}eview,'' \emph{Proc. IEEE}, vol. 107, no.~7, pp. 1256--1301,
  July 2019.

\bibitem{5962989}
M.~S. {Kuran}, H.~B. {Yilmaz}, T.~{Tugcu}, and I.~F. {Akyildiz}, ``{Modulation
  Techniques for Communication via Diffusion in Nanonetworks},'' in \emph{Proc.
  IEEE ICC}, June 2011, pp. 1--5.

\bibitem{7397863}
B.~{Koo}, C.~{Lee}, H.~B. {Yilmaz}, N.~{Farsad}, A.~{Eckford}, and C.~{Chae},
  ``{Molecular {MIMO}: From Theory to Prototype},'' \emph{IEEE J. Sel. Areas
  Commun.}, vol.~34, no.~3, pp. 600--614, March 2016.

\bibitem{6807659}
H.~B. {Yilmaz}, A.~C. {Heren}, T.~{Tugcu}, and C.~{Chae}, ``{Three-Dimensional
  Channel Characteristics for Molecular Communications With an Absorbing
  Receiver},'' \emph{IEEE Commun. Lett.}, vol.~18, no.~6, pp. 929--932, June
  2014.

\bibitem{twoway}
J.~W. {Kwak}, H.~B. {Yilmaz}, N.~{Farsad}, C.~{Chae}, and A.~{Goldsmith},
  ``{Two-Way Molecular Communications},'' \emph{IEEE Trans. Commun.}, pp. 1--1,
  2020.

\bibitem{Yansha16}
Y.~{Deng}, A.~{Noel}, M.~{Elkashlan}, A.~{Nallanathan}, and K.~C. {Cheung},
  ``{Modeling and Simulation of Molecular Communication Systems With a
  Reversible Adsorption Receiver},'' \emph{IEEE Trans. Mol. Biol. Multi-Scale
  Commun.}, vol.~1, no.~4, pp. 347--362, December 2015.

\bibitem{8030318}
Y.~{Deng}, A.~{Noel}, W.~{Guo}, A.~{Nallanathan}, and M.~{Elkashlan},
  ``{Analyzing Large-Scale Multiuser Molecular Communication via 3-D Stochastic
  Geometry},'' \emph{IEEE Trans. Mol. Biol. Multi-Scale Commun.}, vol.~3,
  no.~2, pp. 118--133, Jun. 2017.

\bibitem{6712164}
A.~{Noel}, K.~C. {Cheung}, and R.~{Schober}, ``{Improving Receiver Performance
  of Diffusive Molecular Communication With Enzymes},'' \emph{IEEE Trans.
  Nanobiosci.}, vol.~13, no.~1, pp. 31--43, March 2014.

\bibitem{8922790}
B.~{Li}, W.~{Guo}, X.~{Wang}, Y.~{Deng}, Y.~{Lan}, C.~{Zhao}, and
  A.~{Nallanathan}, ``{CSI-Independent Non-Linear Signal Detection in Molecular
  Communications},'' \emph{IEEE Trans. Signal Process.}, vol.~68, pp. 97--112,
  Dec. 2019.

\bibitem{giannoukos2018chemical}
S.~Giannoukos, D.~T. McGuiness, A.~Marshall, J.~Smith, and S.~Taylor, ``A
  chemical alphabet for macromolecular communications,'' \emph{Anal Chem},
  vol.~90, no.~12, pp. 7739--7746, May 2018.

\bibitem{8489889}
D.~T. {McGuiness}, S.~{Giannoukos}, A.~{Marshall}, and S.~{Taylor},
  ``{Experimental Results on the Open-Air Transmission of Macro-Molecular
  Communication Using Membrane Inlet Mass Spectrometry},'' \emph{IEEE Commun.
  Lett.}, vol.~22, no.~12, pp. 2567--2570, Dec 2018.

\bibitem{farsad2013tabletop}
N.~Farsad, W.~Guo, and A.~W. Eckford, ``Tabletop molecular communication: Text
  messages through chemical signals,'' \emph{PLoS One}, vol.~8, no.~12, p.
  e82935, December 2013.

\bibitem{8924625}
L.~{Grebenstein}, J.~{Kirchner}, W.~{Wicke}, A.~{Ahmadzadeh}, V.~{Jamali},
  G.~{Fischer}, R.~{Weigel}, A.~{Burkovski}, and R.~{Schober}, ``{A Molecular
  Communication Testbed Based on Proton Pumping Bacteria: Methods and Data},''
  \emph{IEEE Trans. Mol. Biol. Multi-Scale Commun.}, vol.~5, no.~1, pp. 56--62,
  Oct 2019.

\bibitem{akyildiz2008nanonetworks}
I.~F. Akyildiz, F.~Brunetti, and C.~Bl{\'a}zquez, ``{Nanonetworks: A New
  Communication Paradigm},'' \emph{Comput. Networks}, vol.~52, no.~12, pp.
  2260--2279, Apr. 2008.

\bibitem{andreescu2004trends}
S.~Andreescu and O.~A. Sadik, ``{Trends and Challenges in Biochemical Sensors
  for Clinical and Environmental Monitoring},'' \emph{Pure Appl Chem}, vol.~76,
  no.~4, pp. 861--878, Jan. 2004.

\bibitem{Alberts2009}
B.~Alberts, D.~Bray, K.~Hopkins, A.~Johnson, J.~Lewis, M.~Raff, K.~Roberts, ,
  and P.~Walter, ``{Essential Cell Biology},'' \emph{(3 ed.). Garl. Press New
  York.}, 2009.

\bibitem{wang2014rapid}
B.~Wang and M.~Buck, ``{Rapid Engineering of Versatile Molecular Logic Gates
  using Heterologous Genetic Transcriptional Modules},'' \emph{Chem. Commun.},
  vol.~50, no.~79, pp. 11\,642--11\,644, Jul. 2014.

\bibitem{Wang11}
B.~Wang, R.~I. Kitney, N.~Joly, and M.~Buck, ``Engineering modular and
  orthogonal genetic logic gates for robust digital-like synthetic biology,''
  \emph{Nat. Commun.}, vol.~2, no. 508, pp. 1--9, Oct. 2011.

\bibitem{kahl2013survey}
L.~J. Kahl and D.~Endy, ``{A Survey of Enabling Technologies in Synthetic
  Biology},'' \emph{J. Biol. Eng.}, vol.~7, no.~1, p.~13, May 2013.

\bibitem{xiang2018scaling}
Y.~Xiang, N.~Dalchau, and B.~Wang, ``{Scaling Up Genetic Circuit Design for
  Cellular Computing: Advances and Prospects},'' \emph{Natural computing},
  vol.~17, no.~4, pp. 833--853, Oct. 2018.

\bibitem{bernard2017synthetic}
E.~Bernard and B.~Wang, ``{Synthetic Cell-based Sensors with Programmed
  Selectivity and Sensitivity},'' in \emph{Biosensors and Biodetection}.\hskip
  1em plus 0.5em minus 0.4em\relax Springer, Mar. 2017, pp. 349--363.

\bibitem{tamsir2011robust}
A.~Tamsir, J.~J. Tabor, and C.~A. Voigt, ``{Robust Multicellular Computing
  using Genetically Encoded NOR Gates and Chemical ‘Wires’},''
  \emph{Nature}, vol. 469, no. 7329, pp. 212--215, Jan. 2011.

\bibitem{ellis2009diversity}
T.~Ellis, X.~Wang, and J.~J. Collins, ``{Diversity-based, Model-guided
  Construction of Synthetic Gene Networks with Predicted Functions},''
  \emph{Nat Biotechnol}, vol.~27, no.~5, pp. 465--471, May 2009.

\bibitem{alon2006introduction}
A.~Uri, \emph{{An Introduction to Systems Biology: Design Principles of
  Biological Circuits}}.\hskip 1em plus 0.5em minus 0.4em\relax London, UK:
  Chapman \& Hall, 2006.

\bibitem{8255057}
Y.~{Deng}, M.~{Pierobon}, and A.~{Nallanathan}, ``{A Microfluidic Feed Forward
  Loop Pulse Generator for Molecular Communication},'' in \emph{Proc. IEEE
  GLOBECOM}, Dec 2017, pp. 1--7.

\bibitem{bi2019chemical}
D.~Bi, Y.~Deng, M.~Pierobon, and A.~Nallanathan, ``{Chemical Reactions-Based
  Microfluidic Transmitter and Receiver for Molecular Communication},''
  \emph{arXiv preprint arXiv:1908.03441}, Aug. 2019.

\bibitem{dadimag}
D.~Bi and Y.~Deng, ``{Digital Signal Processing for Molecular Communication via
  Lego-Like Chemical Reactions-Based Microfluidic Circuits},'' \emph{submitted
  to IEEE Commun. Mag.}

\bibitem{whitesides2006origins}
G.~M. Whitesides, ``{The Origins and the Future of Microfluidics},''
  \emph{Nature}, vol. 442, no. 7101, pp. 368--373, July 2006.

\bibitem{bruustheoretical}
H.~Bruus, \emph{Theoretical microfluidics}.\hskip 1em plus 0.5em minus
  0.4em\relax London, UK: Oxford Univ. Press, 2008.

\bibitem{oh2012design}
K.~W. Oh, K.~Lee, B.~Ahn, and E.~P. Furlani, ``Design of pressure-driven
  microfluidic networks using electric circuit analogy,'' \emph{Lab. Chip},
  vol.~12, no.~3, pp. 515--545, Nov. 2012.

\bibitem{toh2014engineering}
A.~G. Toh, Z.~Wang, C.~Yang, and N.-T. Nguyen, ``{Engineering Microfluidic
  Concentration Gradient Generators for Biological Applications},''
  \emph{Microfluid Nanofluid}, vol.~16, no. 1-2, pp. 1--18, Jul. 2014.

\bibitem{wicke2018modeling}
W.~{Wicke}, T.~{Schwering}, A.~{Ahmadzadeh}, V.~{Jamali}, A.~{Noel}, and
  R.~{Schober}, ``{Modeling Duct Flow for Molecular Communication},'' in
  \emph{Proc. IEEE GLOBECOM}, Dec 2018, pp. 206--212.

\bibitem{bicen2014end}
A.~O. Bicen and I.~F. Akyildiz, ``{End-to-End Propagation Noise and Memory
  Analysis for Molecular Communication over Microfluidic Channels},''
  \emph{IEEE Trans. Commun.}, vol.~62, no.~7, pp. 2432--2443, July 2014.

\bibitem{chang2005physical}
R.~Chang, \emph{{Physical Chemistry for the Biosciences}}.\hskip 1em plus 0.5em
  minus 0.4em\relax Herndon, VA, USA: University Science Books, 2005.

\bibitem{scalise2014designing}
D.~Scalise and R.~Schulman, ``{Designing Modular Reaction-Diffusion Programs
  for Complex Pattern Formation},'' \emph{Technology}, vol.~2, no.~01, pp.
  55--66, Mar. 2014.

\bibitem{harris2010digital}
D.~Harris and S.~Harris, \emph{{Digital Design and Computer
  Architecture}}.\hskip 1em plus 0.5em minus 0.4em\relax CA, San Mateo: Morgan
  Kaufmann, 2010.

\end{thebibliography}

%
%

\end{document}